\documentclass[11pt,envcountsame]{llncs}
\usepackage{verbatim,url,enumerate,color,paralist}
\usepackage{amsmath,amsfonts}
\usepackage{algorithm}
\usepackage{fullpage}

\newlength{\subtextwidth}
\newcommand{\LL}{\mathbb{L}}
\newcommand{\A}{\mathcal{A}}
\newcommand{\I}{\mathcal{I}}

\newcommand{\C}{\mathbb{C}}
\newcommand{\af}{\$}
\newcommand{\D}{\Delta}

\title{Classified Stable Matching}
\author{Chien-Chung Huang\thanks{Research supported by an Alexander von Humboldt fellowship}}
\institute{Max-Planck-Institut f\"ur Informatik, Saarbr\"ucken, 66123, Germany \\
villars@mpi-inf.mpg.de}

\begin{document}
\date{}
\maketitle

\begin{abstract} 
We introduce the {\sc classified stable matching} problem, 
a problem motivated by academic hiring. Suppose that a number of institutes 
are hiring faculty members from a pool of applicants. Both 
institutes and applicants have preferences over the other side. 
An institute classifies the applicants based on their research 
areas (or any other criterion), and, for each class, it sets a lower bound and an upper bound 
on the number of applicants it would hire in that class. 
The objective is to find a stable matching from which 
no group of participants has reason to deviate. Moreover, the matching 
should respect the upper/lower bounds of the classes. 

In the first part of the paper, 
we study classified stable matching problems whose classifications 
belong to a fixed set of ``order types.'' We show that 
if the set consists entirely of downward forests, 
there is a polynomial-time algorithm; otherwise, it is NP-complete 
to decide the existence of a stable matching. 

In the second part, we investigate the problem using a polyhedral approach. Suppose that all classifications are laminar families and there is no lower bound. We propose a set of linear inequalities to describe stable matching polytope and prove that it is integral. 
This integrality allows us to find various optimal stable matchings using Ellipsoid algorithm. By studying the geometric structure of fractional stable matchings, we are able to generalize a theorem of Teo and Sethuraman in the context of classified stable matching: given any number of stable matchings, if every applicant is assigned to his median choice among all stable matchings, the outcome is still a stable matching. Finally, a ramification of our result is the description of the stable matching polytope for the many-to-many (unclassified) stable matching problem. This answers an open question posed by Sethuraman, Teo and Qian. 

\end{abstract}

\newpage

\section{Introduction} 

Imagine that a number of institutes are recruiting faculty members from a pool of applicants. Both sides
have their preferences. It would be ideal if there is a matching from which no applicant
and institute have reason to deviate. If an applicant prefers another institute to
the one he is assigned to (or maybe he is unassigned) and this institute
also prefers him to any one of its assigned applicants, then this institute-applicant pair
is a \emph{blocking pair}. A matching is \emph{stable} if there is no blocking pair.

The above scenario is the well-studied
{\sc hospitals/residents} problem~\cite{Gale:1962,Gusfield:1989}
in a different guise. It is known that stable
matchings always exist and can be found efficiently by the Gale-Shapley algorithm.
However, real world situations can be more complicated.  An institute may have its own
hiring policy and 
may find certain sets of applicants \emph{together} unacceptable. For example,
an institute may have reasons to avoid hiring too many applicants graduated from the same school; or it may want to diversify its faculty so that it can have researchers in many different fields.

This concern motivates us to consider the following problem. 
An institute, besides
giving its preference among the applicants, also classifies
them based on their expertise (or some other criterion).
For each class, it sets an upper bound and a lower bound on the number 
of applicants it would hire. Each institute defines its own classes and classifies the
applicants in its own way (and the classes need not be disjoint). 
We consider this flexibility a desirable feature, as there are
some research fields whose boundaries are blurred; moreover, some versatile researchers
may be hard to categorize.

We call the above problem {\sc classified stable matching}. 
Even though motivated by academic hiring, 
it comes up any time objects on one side of the matching
have multiple partners that may be classified. For example, the two sides can be
jobs and machines;
each machine is assigned several jobs but perhaps cannot take two jobs with heavy memory requirements.

To make the problem precise, 
we introduce necessary notation and terminology. A set $\A$ of applicants and 
a set $\I$ of institutes are given. 
Each applicant/institute has a strictly-ordered (but not necessarily
complete) preference list over the other side. The notation $\succeq_{x}$ 
indicates either strictly better or equal
in terms of preference of an entity $e \in \A \cup \I$ while $\succ_{e}$ means strictly better.
For example, if applicant $a \in \A$ strictly prefers institute $i\in \I$ to another
institute $i' \in \I$, we write $ i \succ_{a} i'$.
The preference list of institute $i$ is denoted as $\LL^i$. The set of applicants on $\LL^i$ who rank higher (respectively lower) than some particular applicant $a$ are written as $\LL^i_{\succ a}$ (respectively $\LL^i_{\prec a}$). 

An institute $i$ has a \emph{capacity} $Q(i) \in \mathbb{Z}^{+}$, the maximum
number of applicants it can hire. It defines its own classification $\mathbb{C}(i)=
\{C^i_j\}_{j=1}^{|\C(i)|}$, which is
a family of sets over the applicants in its preference list.
Each class $C^i_j \in \mathbb{C}(i)$ has an upperbound $q^{+}(C^i_j) 
\in \mathbb{Z}^{+}$ and a lowerbound $q^{-}(C^i_j) \in \mathbb{Z}^{+}\cup \{0\}$, 
on the number of applicants it would hire in that class. 
Given a matching $\mu$, $\mu(a)$ is the institute applicant $a$ is assigned
to. We write $\mu(i) = (a_{i1},a_{i2},\cdots,a_{ik})$, $ k\leq Q(i)$
to denote the set of applicants institute $i$ gets in $\mu$, where $a_{ij}$ are listed in decreasing order based on its preference list. In this paper, we will slightly abuse notation, treating
an (ordered) tuple such as $\mu(i)$ as a set. 

\begin{definition} Given a tuple $t=(a_{i1},a_{i2},\cdots, a_{ik})$ where $a_{ij}$ are ordered based on their decreasing rankings on institute $i$'s preference list, it is said to be a feasible tuple of institute $i$, or just feasible for short, if the following conditions hold: 

\begin{itemize}
\item $k \leq Q(i)$;
\item given any class $C^i_j \in \C(i)$, $q^{-}(C^i_j) \leq |t \cap C^i_j| \leq q^{+}(C^i_j)$. 

\end{itemize}
\end{definition}

\begin{definition} A matching $\mu$ is feasible if all the tuples $\mu(i)$, $i\in \I$ are feasible. 
A feasible matching is stable if and only if there is no \emph{blocking group}.
A blocking group is defined as follows. Let
$\mu(i) = (a_{i1},a_{i2},\cdots, a_{ik}), k\leq Q(i)$.
A feasible tuple $g = (a'_{i1},a'_{i2},\cdots,a'_{ik'}), k \leq k'
\leq Q(i)$, forms a blocking group $(i \mbox{;} g)$ with institute $i$ if
\begin{itemize}
\item for $1 \leq j \leq k, i \succeq_{a'_{ij}} \mu(a'_{ij})$ and 
$a'_{ij} \succeq_{i} a_{ij}$;
\item either there exists $l, 1 \leq l \leq k$ such that
$a'_{il} \succ_{i} a_{il}$ and $i \succ_{a'_{il}} \mu(a'_{il})$, or that $k' > k$.
\end{itemize}
\label{def:first}
\end{definition}
Informally speaking, the definition requires that for a blocking group
to be formed, all involved applicants have to be willing to switch to,
or stay with, institute $i$.
The collection of applicants in the blocking group should still 
respect  the upper and lower bounds in each class; moreover, 
the institute gets a \emph{strictly better deal} (in the
Pareto-optimal sense). 
Note that when there is no class lower bound, then the stable matching as defined in Definition~\ref{def:first} can be equivalently defined as a feasible matching without the conventional \emph{blocking pairs} (see Lemma~\ref{pro:blockingPair} 
in Section~\ref{sec:polyhedral}). When the class lower bound is present, the definition of the blocking groups captures our intuition that an institute should not indiscriminately replace a lower ranking applicant assigned to it with a higher applicant (with whom it forms a blocking pair), otherwise, the outcome for it may not be a feasible one. 

In our proofs, we often use the notation
$\mu(i)|^{a}a'$
 to denote a tuple formed by replacing $a \in \mu(i)$ with $a'$. The order of
the tuple $\mu(i)|^{a}a'$ is still based on institute $i$'s preference list. If we write $\mu(i)|a$,
then this new tuple is obtained by adding $a$ into $\mu(i)$ and re-ordered.
In a matching $\mu$, if a class $C^i_j$ is fully-booked, i.e. $|\mu(i)\cap C^i_j| = 
q^{+}(C^i_j)$, we often refer to such a class as a ``bottleneck'' class. We also define an ``absorption'' operation: given a set $B$ of classes, $\Re(B)$ returns the set of classes which are not entirely contained in other classes in $B$.

\subsubsection*{Our Results}

It would be of interest to know how complicated the classifications 
of the institutes can be while still allowing the problem 
a polynomial time algorithm. 
In this work, we study the {\sc classified stable matching} 
problems whose classifications
belong to a fixed set of ``order types.'' The order type of a classification 
is the inclusion poset of all non-empty intersections 
of classes. We introduce necessary definitions to make our statement precise. 

\begin{definition} The \emph{class inclusion poset} 
$P(i) = (\overline{\C}(i), \preceq)$ of an institute $i$ is composed of
sets of the elements from $\LL^i$: $\overline{\C}(i)= \{\overline{C}| 
\overline{C} = C^i_j \cap C^i_k, \mbox{ where } C^i_j, 
C^i_k \in \C(i)\}$\footnote{Note that this definition allows a class 
to intersect itself, i.e., $\overline{C}=  C^i_j \cap C^i_j$. This implies that 
$\overline{\C}(i) \supseteq \C(i)$.}. 
In $P(i)$, $\overline{C}^i_j \succ \overline{C}^i_k$
if $\overline{C}^i_j \supset \overline{C}^i_k$; and  $\overline{C}^i_j \| \overline{C}^i_k$
if $\overline{C}^i_j \not \supset \overline{C}^i_k$ and $\overline{C}^i_k \not \supset \overline{C}^i_j$.

\label{def:inclusion_poset}
\end{definition}

\begin{definition} Let $\mathbb{P} = \{P_1, P_2,\cdots, P_k\}$ 
be a set of posets.
A classified stable matching instance $(\A,\I)$ belongs to 
the group of $\mathbb{P}$-{\sc classified stable matching} problems
if for each poset $P_j \in \mathbb{P}$, there exists an institute $i \in \I$
whose class inclusion poset $P(i)$ is isomorphic to $P_j$ and conversely, 
every class inclusion poset $P(i)$ is  isomorphic to a poset in $\mathbb{P}$.
\label{def:posetDefinition}
\end{definition}

We call a poset a \emph{downward forest} if given any element, no two of its successors are incomparable. Our first main result is the following dichotomy theorem.

\begin{theorem} Let $\mathbb{P}  = \{P_1, P_2,\cdots, P_k\}$ 
be a set of posets. $\mathbb{P}$-classified stable matching problems 
can be solved in polynomial time if every poset 
$P_j \in \mathbb{P}$ is a downward forest; on the other hand, if
$\mathbb{P}$ contains a poset $P_j$ which is not a downward forest, 
the existence of a stable matching is NP-complete.
\label{thm:dichotomy}
\end{theorem}

We remark that if $\mathbb{P}$ is entirely composed of downward forests, then every 
classification $\C(i)$ must be a \emph{laminar} family\footnote{A laminar family $\mathcal{F}$ 
has no pair of intersecting classes, that is, if $A, B \in \mathcal{F}$, then either 
$A\cap B = \emptyset$, or $A \subseteq B$, or $B \subseteq A$.}.
In this case, we call the problem {\sc laminar classified stable matching} (henceforth $\textbf{LCSM}$). 

We present an $O(m^2)$ time algorithm for \textbf{LCSM}, where $m$ is the total size of all preferences. Our algorithm is extended from the Gale-Shapley algorithm. Though intuitive, its correctness is difficult to argue due to various constraints\footnote{If there is no lower bound on the classes, the proof can be significantly simplified.}. Furthermore, we show that several well-known structural results in the {\sc hospitals/residents} problem can be further generalized in \textbf{LCSM}. 
On the other hand, if some institute $i$ has a classification $\C(i)$ violating 
laminarity, then $\mathbb{P}$ must contain a poset which has a ``V'' (where the 
``bottom'' is induced by two intersecting classes in $\C(i)$ which are its parents 
``on top.'') We will make use of this fact to design a gadget for our NP-complete reduction. 
In particular, in our reduction, all institutes only use upperbound constraints. 
Sections~2 and~3 will be devoted to these results. 

Our dichotomy theorem implies a certain limit 
on the freedom of the classifications defined by the institutes.
For example, an institute may want to classify the applicants based on two different criteria 
simultaneously (say by research fields and gender); however, our result implies this may cause the problem to become intractable. 

In the second part, we study \textbf{LCSM} using a mathematical programming approach. Assume that there is no lower bound on the classes. We extend the set of linear inequalities used by Ba\"{i}ou and Balinski~\cite{Baiou:2000} to describe stable matchings and generalize a bin-packing algorithm of Sethuraman, Teo, and Qian~\cite{Sethuraman:2006a} to prove that the polytope is integral. The integrality of our polytope allows us to use suitable objective functions to obtain various optimal stable matchings using Ellipsoid algorithm. As our LP has an exponential number of constraints, we also design a separation oracle. 

By studying the geometric structure of fractional stable matchings, we are able to generalize a theorem of Teo and Sethuraman~\cite{Teo:2001a}: in (one-to-one) stable marriage, given any number of stable matchings, if we assign every man his median choice among all women with whom he is matched in the given set of matchings and we do similarly for women, the outcome is still a stable matching. This theorem has been generalized in the context of {\sc hospitals/residents} problem~\cite{Fleiner:2002,Klaus:2006,Sethuraman:2006a}. We prove that in \textbf{LCSM}, this theorem still holds: if we apply this ``median choice operation'' on all applicants, the outcome is still a stable matching\footnote{However, this ``median choice operation'' on the applicants' side does not mean that all institutes will get their (lexicographically) median outcome. Thus, the ``fairness'' of the resultant matching is to a certain degree lost in our generalizations. See Section~4 for details.}. 

A final ramification of our polyhedral result is an answer to an open question posed by Sethuraman, Teo and Qian~\cite{Sethuraman:2006a}: how do we describe the stable matching polytope in the classical ``unclassified'' many-to-many stable matching problem? We show this problem can be reduced to \textbf{LCSM} by suitable cloning and classifications. 

All the polyhedral results will be presented in Section~4. In Section~5 we conclude. Omitted proofs and details can be found in the appendix. 

\subsection{Related Work}

Stable matching problems have drawn the intensive attention of researchers in various disciplines in the past decades since the seminal paper of Gale and Shapley~\cite{Gale:1962}. For a summary, see~\cite{Gusfield:1989,Knuth:1976,Roth:1990}. Vande Vate~\cite{Vande:1989} initiated the study of stable matching using mathematical programming approach; further developments using this approach can be found 
in~\cite{Abeledo:1994,Baiou:2000,Roth:1993,Rothblum:1992,Sethuraman:2001,Sethuraman:2006a,Teo:2001a}.  

Fleiner~\cite{Fleiner:2003} studied the many-many stable matching in a much more general context. Using a fixed-point approach, he proved that stable matchings always exist provided that the preference of each entity is a \emph{substitutable choice function}. Roughly speaking, such a function can be realized by imposing a matroid over a linear order of elements. In \textbf{LCSM}, supposing that there is no lower bound on the classes, then each laminar family  is equivalent to a partition matroid. We prove that stable matchings always exist in this situation. Hence, our algorithm in Section~2 can be seen as a constructive proof of a special case of Fleiner's existence theorem. 

Abraham, Irving and Manlove introduced the {\sc student-project  allocation} problem~\cite{Abraham:2007a}. It can be shown that in \textbf{LCSM}, if all classifications are just partitions over the applicants and there is no lower bound, our problem is equivalent to a special case of their problem. They posed the open question whether there is a polynomial time algorithm for their problem if there is lower bound on the projects (classes). Our result in Section~2 gives a partial positive answer. 

Two recent works~\cite{Biro:2009,Hamada:2008} also consider the Hospitals/Residents problem in the context of having lower bounds on the hospitals' side. In~\cite{Biro:2009}, an interesting variation of the Hospitals/Residents problem which has a similar flavor to the current work is defined as follows: each hospital has its individual quota and sets of hospitals may also have collective quotas.


\section{An Algorithm for Laminar Classified Stable Matching} 

In this section, we present a polynomial time algorithm to find a stable matching
 if it exists in the given \textbf{LCSM} instance, otherwise, to report that none exists.

We pre-process our instance as follows. If applicant $a$ is on institute $i$'s preference list, we add a class $C^i_{a1}=\{a\}$ into $\C(i)$. Furthermore, we also add a class $C^i_{\sharp}$ into $\C(i)$ including all applicants in $\LL^i$. After this pre-processing, the set of classes in $\C(i)$ form a tree whose root is the $C^i_{\sharp}$; moreover, an applicant $a$ belongs to a sequence of classes $a(\C(i))=(C^i_{a1}, C^i_{a2},\cdots, C^i_{az}(=C^i_{\sharp}))$, which forms a path from the leaf to the root in the tree (i.e., $C^i_{aj}$ is a super class of $C^i_{aj'}$, provided $j' <j$.) For each non-leaf class $C^i_j$, let $c(C^i_j)$ denote the set of its child classes in the tree. We can assume without loss of generality that $q^{-}(C^i_j)  \geq \sum_{C^i_k \in c(C^i_j)}q^{-}(C^i_k)$ for any non-leaf class $C^i_j$. Finally, let 
$q^{+}(C^i_{\sharp}) := Q(i)$, $q^{-}(C^i_{\sharp}) := \sum_{C^i_k \in c(C^i_{\sharp})} q^{-}(C^i_k)$; for all applicants $a \in \LL^i$, $q^{+}(C^i_{a1}) :=1$ and $q^{-}(C^i_{a1}):=0$.

Our algorithm finds an \emph{applicant-optimal-institute-pessimal} stable matching. The applicant-optimality means that all applicants get the best outcome among all stable matchings; on the other hand, institute-pessimality means that all institutes get an outcome which is ``lexicographically'' the worst for them. To be precise, suppose that $\mu(i)=(a_{i1},a_{i2},\cdots, a_{ik})$ and $\mu'(i)=(a_{i1},a_{i2},\cdots, a_{ik})$ are the outcomes of two stable matchings for institute $i$\footnote{In \textbf{LCSM}, an institute always gets the same number of applicants in all stable matchings. See Theorem~\ref{the:strongerRuralHospital} below.}. If there exists $k' \leq k$ so that 
$a_{ij} =a'_{ij}$, for all $1 \leq j \leq k'-1$ and $a_{ik'} \succ_{i} a'_{ik'}$, then institute $i$ is lexicographically better off in $\mu$ than in $\mu'$.

We now sketch the high-level idea of our algorithm. We let applicants ``propose'' to the institutes 
from the top of their preference lists. Institutes make the decision of acceptance/rejection of the proposals based on certain rules (to be explained shortly). Applicants, if rejected, propose to the next highest-ranking institutes on their lists. The algorithm terminates when all applicants either end up with some institutes, or run out of their lists. Then we check whether the final outcome meets the upper and lower bounds of all classes. If yes, the outcome is a stable matching; if no, there is no stable matching in the given instance. 

How the institutes make the acceptance/rejection decisions is the core of our algorithm. Intuitively, when an institute gets a proposal, it should consider two things: \textrm{(i)} will adding this new applicant violate the upper bound of some class? \textrm{(ii)} will adding this applicant deprive other classes of their necessary minimum requirement? If the answer to any of the two questions is positive, the institute should not just take the new applicant unconditionally; instead, it has to reject someone it currently has (not necessarily the new comer). 

Below we will design two invariants for all classes of an institute. Suppose that institute $i$ gets a proposal from applicant $a$, who belongs to a sequence of classes $a(\C(i)) = (C^i_{a1},C^i_{a2},\cdots, C^i_{\sharp})$. We check this sequence of classes from the leave to the root. If adding applicant $a$ into class $C^i_{aj}$ does not violate these invariants, we climb up and see if adding applicant $a$ into $C^i_{a(j+1)}$ violates the invariant. If we can reach all the way to $C^i_{\sharp}$ without violating the invariants in any class in $a(\C(i))$, applicant $a$ is just added into institute $i$'s new collection. If, on the other hand, adding applicant $a$ into 
$C^i_{a(j+1)}$ violates the invariants, institute $i$ rejects \emph{some} applicant in $C^i_{a(j+1)}$ who is from \emph{a sequence of subclasses of $C^i_{a(j+1)}$ which can afford to lose one applicant}. 

We define a \emph{deficiency} number $\D(C^i_j)$ for each class $C^i_j \in \C(i)$. Intuitively, 
the deficiency number indicates how many more applicants are necessary for class $C^i_j$ to meet the lower bound of all its subclasses. This intuition translates into the following invariant: 
\begin{quote} \textbf{Invariant~$A$}: $\D(C^i_j) \geq \sum_{C^i_k \in c(C^i_j)}\D(C^i_k), \forall C^i_j \in \C(i), c(C^i_j) \neq \emptyset, \forall i \in \I$.
\end{quote}

In the beginning, $\D(C^i_j)$ is set to $q^{-}(C^i_j)$ and we will explain how $\D(C^i_j)$ is updated shortly. Its main purpose is to make sure that after adding some applicants into $C^i_j$, there is still enough ``space'' for other applicants to be added into $C^i_j$ so that we can satisfy the lower bound of all subclasses of $C^i_j$. In particular, we maintain

\begin{quote} \textbf{Invariant~$B$}: $q^{-}(C^i_j) \leq |\mu(i) \cap C^i_j| + \D(C^i_j) \leq q^{+}(C^i_j), \forall C^i_j \in \C(i), \forall i \in \I$.
\end{quote}

We now explain how $\Delta(C^i_j)$ is updated. Under normal circumstances, we decrease $\D(C^i_j)$ by 1 once we add a new applicant into $C^i_j$. However, if Invariant~$A$ is already ``tight'', i.e.,  $\D(C^i_j) = \sum_{C^i_k \in c(C^i_j)}\D(C^i_k)$, then we add the new applicant $C^i_j$ without decreasing $\D(C^i_j)$. The same situation may repeat until the point that $|\mu(i) \cap C^i_j| + \D(C^i_j) = q^{+}(C^i_j)$ and adding another new applicant in $C^i_j$ is about to violate Invariant~$B$. In this case, something has to be done to ensure that Invariant~$B$ holds: some applicant in $C^i_j$ has to be rejected, and the question is whom?

Let us call a class a \emph{surplus} class if $|\mu(i) \cap C^i_j| + \D(C^i_j) > q^{-}(C^i_j)$ and we define an \emph{affluent set} for each class $C^i_j$ as follows: 

$$\af(C^i_{j},\mu(i)) =  \{a|a \in \mu(i) \cap C^i_{j}; 
\mbox{ for each } C^i_{j'} \in a(\C(i)) \mbox{ and } C^i_{j'} \subset C^i_{j}, 
|\mu(C^i_{j'})| +  \Delta(C^i_{j'})  > q^{-}(C^i_{j'})\}.$$ 

In words, the affluent set $\af(C^i_j,\mu(i))$ is composed of the set of applicants currently assigned to institute $i$, part of $C^i_j$, and each of whom belonging to a sequence of surplus subclasses of $C^i_j$. In our algorithm, to prevent Invariant~$B$ from being violated in a non-leaf class $C^i_j$, institute $i$ rejects the lowest ranking applicant $a$ in the affluent set $\af(C^i_j,\mu(i))$. 

The pseudo-code of the algorithm is presented in Figure~\ref{fig:AlgA}.

\vspace*{-0.05in}\hspace*{-0.4in}
\begin{figure*}[h]\footnotesize
\hrule
\begin{tabbing}
\hspace{15pt}\=\hspace{20pt}\=\hspace{11pt}\=
\hspace{11pt}\=\hspace{11pt}\=\hspace{11pt}\=\hspace{11pt}\\ \vspace*{-0.45in}

\textrm{Initialization} \\

\>0: \> $\forall i\in \I$, $\forall C^i_j \in \C(i)$, $\D(C^i_j) := q^{-}(C^i_j)$;\\

\textrm{Algorithm} \\
\>1: \> \textbf{While} there exists an applicant $a$ unassigned and he has not been rejected by all institutes on his list \\
\>2: \> \> Applicant $a$ proposes to the highest ranking institute $i$ to whom he has not proposed so far; \\
\>3: \> \> Assume that $a(\C(i)) = (C^i_{a1},C^i_{a2},\cdots, C^i_{az}(=C^i_{\sharp}))$; \\
\>4: \> \> $\mu(i) := \mu(i) \cup \{a\}$ // Institute $i$ accepts applicant $a$ provisionally; \\
\>5: \> \> \textbf{For} $t=2$ \textbf{To} $z$  // applicant $a$ can be added into $C^i_{a1}$ directly; \\
\>6: \> \> \> \textbf{If}  $\D(C^i_{at}) > \sum_{C^i_k \in c(C^i_{at})} \D(C^i_{k})$
\textbf{ Then } $\D(C^i_{at}):= \D(C^i_{at})-1$; \\
\>7: \> \> \> \textbf{If} $\#(C^i_{at}) + \D(C^i_{at}) > q^{+}(C^i_{at})$ \textbf{Then} \\
\>8 \> \> \> \> Let $\af(C^i_{at},\mu(i)) =  \{a|a \in \mu(i) \cap C^i_{at}; 
\mbox{ for each } C^i_{j'} \in a(\C(i)) \mbox{ and } C^i_{j'} \subset C^i_{at}, 
|\mu(C^i_{j'})| +  \Delta(C^i_{j'})  > q^{-}(C^i_{j'})\}$; \\
\> 9 \> \> \> \> Let the lowest ranking applicant in $\af(C^i_{at},\mu(i))$ be $a^{\dagger}$; \\
\> 10 \> \> \> \> $\mu(i) := \mu(i) \backslash \{a^{\dagger}\}$ // Institute $i$ rejects applicant $a^{\dagger}$; \\
\>11: \> \> \> \> \textbf{GOTO 1};\\
\>12: \> \textbf{If} there exists an institute $i$ with $\D(C^i_{\sharp}) >0$ \textbf{Then} Report "There is no stable matching"; \\
\>13: \> \textbf{Else} Return the outcome $\mu$, which is a stable matching;

\end{tabbing}
\vspace*{-0.15in}
\caption{The pseudo code of the algorithm. It outputs the applicant-optimal-institute-pessimal matching $\mu$ if it exists; otherwise, it reports that there is no stable matching.}
\hrule
\label{fig:AlgA}
\end{figure*}

\subsection{Correctness of the Algorithm} 

In our discussion, $C^i_{at}$ is a class in $a(\C(i))$, where $t$ is the index based on the size of the class $C^i_{at}$ in $a(\C(i))$. Assume that during the execution of the algorithm, applicant $a$ proposes to institute $i$ and when the index $t$ of the \textbf{For} loop of Line 5 becomes $l$ and results in $a^{\dagger}$ being rejected, 
we say applicant $a$ is \emph{stopped} at class $C^i_{al}$, and class $C^i_{al}$ \emph{causes} applicant $a^{\dagger}$ to be rejected. 

The first lemma describes some basic behavior of our algorithm. 

\begin{lemma}

\begin{enumerate}

\item[(i)] Immediately before the end of the while loop, Invariants~$A$ and $B$ hold.

\item[(ii)] Let applicant $a$ be the new proposer and assume he is stopped at class $C^i_{al}$. Then

\begin{enumerate}
\item[(iia)] Between the time interval that he makes the new proposal and 
he is stopped at $C^i_{al}$, $\D(C^i_{at})$ remains unchanged, for all $1 \leq t \leq l$;
moreover, given any class $C^i_{at}$, $2 \leq t \leq l$, $\D(C^i_{at})= \sum_{C^i_k \in c(C^i_{at})}\D(C^i_k)$.  

\item[(iib)] When $a$ is stopped at a non-leaf class $C^i_{al}$, $\$(C^i_{al},\mu(i)) \neq \emptyset$; in particular, any class $C^i_{at}$, $1 \leq t \leq l-1$, is temporarily a surplus class. 
\end{enumerate}

\item[(iii)]
 Immediately before the end of the while loop, if class $C^i_j$ is a non-leaf surplus class, 
then $\D(C^i_j) = \sum_{C^i_k \in c(C^i_j)}
 \D(C^i_k)$.

\item[(iv)] Suppose that applicant $a$ is the new proposer and $C^i_{al} \in a(\C(i))$ causes applicant $a^{\dagger}$ to be rejected and 
$a^{\dagger}(\C(i)) = (C^i_{a^{\dagger}1}, C^i_{a^{\dagger}2}, \cdots,  C^i_{a^{\dagger}l^{\dagger}}(=C^i_{al}),\cdots)$. Then immediately before the end of the while loop, $\D(C^i_{a^{\dagger}t'}) 
= \sum_{C^i_k \in c(C^i_{a^{\dagger}t'})} \D(C^i_k)$, for all $2 \leq t' \leq l^{\dagger}$;  moreover, $|\mu(i) \cap C^i_{a^{\dagger}l^{\dagger}}| +  \D(C^i_{a^{\dagger}l^{\dagger}}) = q^{+}(C^i_{a^{\dagger}l^{\dagger}}).$ 

\end{enumerate}
\label{lem:Invariant}
\end{lemma}

\begin{proof} \textrm{(i)} can be proved by induction on the number of proposals institute $i$ gets. For \textrm{(iia)}, since Invariant~$A$ is maintained, if $\D(C^i_{at})$ is decreased for some class $C^i_{at}$, $1 \leq t \leq l$, the algorithm will ensure that applicant $a$ would not be stopped in any class, leading to a contradiction.  Now by \textrm{(iia)}, the set of classes $\{C^i_{at}\}_{t=1}^{l-1}$ are (temporarily) surplus classes when applicant $a$ is stopped at $C^i_{al}$, so $\af(C^i_{al},\mu(i)) \neq \emptyset$, establishing \textrm{(iib)}. Note that this also guarantees that the proposed algorithm is never ``stuck.''  

\textrm{(iii)} can be proved inductively on the number of proposals that institute $i$ gets. Assuming $a$ is the new proposer, there are two cases: (1) Suppose that applicant $a$ is not stopped in any class. Then a class $C^i_{at} \in a(\C(i))$ can become surplus only if the stated condition holds ;  (2) Suppose that applicant $a$ is stopped in some class, which 
causes $a^{\dagger}$ to be rejected. Let the smallest class containing both $a$ and $a^{\dagger}$ be $C^i_{al'}$. Applying \textrm{(iia)} and observing the algorithm, it can be verified that only a class $C^i_{at} \subset C^i_{al'}$ can become a surplus class and for such a class, the stated condition holds. 

Finally, for the first part of \textrm{(iv)}, let $C^i_{al'}$ denote the smallest class containing both $a$ and $a^{\dagger}$.  Given a class $C^i_{a^{\dagger}t'}$, if $C^i_{al'} \subseteq C^i_{a^{\dagger}t'} \subseteq C^i_{al}$, \textrm{(iia)} gives the proof. If $C^i_{a^{\dagger}t'} \subset C^i_{al'}$, observe that the former must have been a surplus class right before applicant $a$ made the new proposal. Moreover, before applicant $a$ proposed, \textrm{(iii)} implies that for a non-leaf class $C^i_{a^{\dagger}t'} \subset C^i_{al'}$, 
the stated condition regarding the deficiency numbers is true. 
The last statement of \textrm{(iv)} is by the algorithm and Invariant~$B$. \qed 
 
\end{proof}

\begin{lemma} Assume that $a^{\dagger}(\C(i)) = (C^i_{a^{\dagger}1}, 
C^i_{a^{\dagger}2}, \cdots, C^i_{a^{\dagger}l^{\dagger}}, \cdots)$. 
During the execution of the algorithm, suppose that class $C^i_{a^{\dagger}l^{\dagger}}$ causes applicant $a^{\dagger}$ to be rejected. In the subsequent execution of the algorithm, assuming that $\mu(i)$ is the assignment of institute $i$ at the end of the while loop, then 
there exists $l^{\ddag}$, where $l^{\ddag} \geq l^{\dagger}$ such that $|\mu(i) \cap  C^i_{a^{\dagger}l^{\ddag}}| + \D(C^i_{a^{\dagger}l^{\ddag}}) = q^{+}(C^i_{a^{\dagger}l^{\ddag}})$; furthermore, for all $2 \leq t \leq l^{\ddag}$, all applicants in $\af(C^i_{a^{\dagger}t}, \mu(i))$ rank higher than $a^{\dagger}$. Moreover, for all $2 \leq t \leq l^{\ddag}$, $\D(C^i_{a^{\dagger}t}) = \sum_{C^i_k \in c(C^i_{a^{\dagger}t})} \D(C^i_k)$. 
\label{lem:gettingBetter}
\end{lemma}

\begin{proof} We prove based on the induction on the number of proposals institute $i$ receives after $a^{\dagger}$ is rejected. The base case is when $a^{\dagger}$ is just rejected. Let $l^{\ddag}= l^{\dagger}$. Then it is obvious that all applicants in the affluent sets 
$\af(C^i_{a^{\dagger}t}, \mu(i))$, $2 \leq t  \leq l^{\ddag}$, rank higher than $a^{\dagger}$ and the rest of the lemma holds by 
Lemma~\ref{lem:Invariant}\textrm{(iv)}. 

For the induction step, let $a$ be the new proposer. There are four cases. Except the second case, we let $l^{\ddag}$ remain unchanged after $a$'s proposal. 

\begin{itemize} 

\item Suppose that $a \not \in C^i_{a^{\dagger}l^{\ddag}}$ and he does not cause anyone in $C^i_{a^{\dagger}l^{\ddag}}$ to be rejected. Then the proof is trivial.

\item Suppose that $a \not \in C^i_{a^{\dagger}l^{\ddag}}$ and he is stopped in class $C^i_{al}$, which causes an applicant $a^{*} \in C^i_{a^{\dagger}l^{\ddag}}$ to be rejected. 
$a^{*}$ must be part of the affluent set $\af(C^i_{a^{\dagger}l^{\ddag}}, \mu(i))$ 
before $a$ proposed. By induction hypothesis, $a^{*} \succ_{i} a^{\dagger}$. Moreover, 
since $a^{*}$ is chosen to be rejected, all the applicants in the (new) affluent sets 
$\af(C^i_{a^{\dagger}t}, \mu(i))$, for each class $C^i_{a^{\dagger}t}$, where 
$ C^i_{a^{\dagger}l^{\ddag}} \subset C^i_{a^{\dagger}t}  \subseteq C^i_{al}$, rank higher than $a^{*}$, hence, also higher than $a^{\dagger}$. Now let $C^i_{al}$ be the new $C^i_{a^{\dagger}l^{\ddag}}$ and the rest of the lemma follows from Lemma~\ref{lem:Invariant}\textrm{(iv)}. 

\item Suppose that $a \in C^i_{a^{\dagger}l^{\ddag}}$ and he is not stopped in $ C^i_{a^{\dagger}l^{\ddag}}$ or any of its subclasses. We argue that $a$ must be accepted without causing anyone to be rejected; moreover, the applicants in all affluent sets
$\af(C^i_{a^{\dagger}t}, \mu(i))$, for all $1 \leq t \leq l^{\ddag}$ remain unchanged. Let the smallest class in $a^{\dagger}(\C(i))$ containing $a$ be $C^i_{a^{\dagger}\tilde{l}}$. Note that before $a$ proposed, the induction hypothesis states that $ 
|\mu(i) \cap C^i_{a^{\dagger}l^{\ddag}}| +\D(C^i_{a^{\dagger}l^{\ddag}}) 
= q^{+}(C^i_{a^{\dagger}l^{\ddag}}) $. As applicant $a$ is not stopped at $C^i_{a^{\dagger}l^{\ddag}}$, the set of values $\D(C^i_{a^{\dagger}t})$, $ \tilde{l} \leq t \leq l^{\ddag}$, must have decreased during his proposal and this implies that he will not be stopped in any class. 

Now let $a(\C(i)) = (C^i_{a1},\cdots, C^i_{al}, C^i_{a(l+1)}(=C^i_{a^{\dagger}\tilde{l}}),\cdots)$. Since $\D(C^i_{a^{\dagger}\tilde{l}}) = \sum_{C^i_k \in c(C^i_{a^{\dagger}\tilde{l}})} \D(C^i_k)$ before applicant $a$ proposed by the induction hypothesis, for $\D(C^i_{a^{\dagger}\tilde{l}})$ to decrease, $\D(C^i_{al})$ must have decreased as well. Choose the smallest class $C^i_{al^{*}} \subset C^i_{a^{\dagger}\tilde{l}}$ whose value $\D(C^i_{al^{*}})$ has decreased during $a$'s proposal. We claim that $C^i_{al^{*}}$ must have been a non-surplus class \emph{before and after} applicant $a$'s proposal. If the claim is true, then all the affluent sets $\af(C^i_{a^{\dagger}t}, \mu(i))$, for all $1 \leq t \leq l^{\ddag}$, remain unchanged after applicant $a$'s proposal. 

It is obvious that $C^i_{al^{*}} \neq C^i_{a1}$. 
So assume that $C^i_{al^{*}}$ is a non-leaf class. 
Suppose for a contradiction that $C^i_{al^{*}}$ was a surplus class before $a$ proposed. 
Lemma~\ref{lem:Invariant}\textrm{(iii)} implies that $\D(C^i_{a^{\dagger}l^{*}}) = \sum_{C^i_k \in c(C^i_{a^{\dagger}l^{*}})} \D(C^i_k)$ before $a$ proposed. Then for 
$\D(C^i_{a^{\dagger}l^{*}}) $ to decrease during $a$'s proposal, $\D(C^i_{a^{\dagger}(l^{*}-1)}) $ must have decreased as well. But then this contradicts our choice of $C^i_{a^{\dagger}l^{*}}$. So we establish that $C^i_{al^{*}}$ was not surplus and remains   so after $a$'s proposal. 

\item Suppose that $a \in C^i_{a^{\dagger}l^{\ddag}}$ and when he reaches a subclass of 
$C^i_{a^{\dagger}l^{\ddag}}$ or the class itself, the latter causes some applicant $a^{*}$ to be rejected. To avoid trivialities, assume $a \neq a^{*}$. Let the smallest class in $a^{\dagger}(\C(i))$ containing $a$ be $C^i_{a^{\dagger}\tilde{l}}$ and the smallest class in $a^{\dagger}(\C(i))$ containing $a^{*}$ be $C^i_{a^{\dagger}l^{*}}$. Below we only argue that the case that $C^i_{a^{\dagger}\tilde{l}} \subseteq C^i_{a^{\dagger}l^{*}}$. The other case that 
$C^i_{a^{\dagger}l^{*}} \subset C^i_{a^{\dagger}\tilde{l}}$ follows essentially the same argument. 

After $a$'s proposal, observe that only the affluent sets $\af(C^i_{a^{\dagger}t}, \mu(i))$, 
$\tilde{l} \leq t < l^{*}$, can have new members (who are from the child class of $C^i_{a^{\dagger}\tilde{l}}$ containing $a$). 
Without loss of generality, let $G$ be the set of new members added into one of the any above sets.   

To complete the proof, we need to show that either $G=\emptyset$ or all members in $G$ rank higher than $a^{\dagger}$.  
If before applicant $a$ proposed, $a^{*}$ belonged to a sequence of surplus classes $C^i_{a^{*}t} \subset C^i_{a^{\dagger}l^{*}}$, he was also part of the affluent set $\af(C^i_{a^{\dagger}l^{*}}, \mu(i))$ or part of $\mu(i) \cap C^i_{a^{\dagger}1}$ before $a$ proposed. By induction hypothesis, $ a^{*} \succ_{i} a^{\dagger}$. Observing Lemma~\ref{lem:Invariant}\textrm{(iib)}, all applicants in $G$ must rank higher than $a^{*}$, hence also than $a^{\dagger}$. On the other hand, if $a^{*}$ belongs to some class $C^i_{a^{*}t} \subset C^i_{a^{\dagger}l^{*}}$ which was not surplus before $a$ proposed, then $C^i_{a^{*}\tilde{l}}=C^i_{a^{*}l^{*}}$ and $C^i_{a^{*}t}$ must also contain $a$ and remain a non-surplus class after $a$'s proposal. In this case $G = \emptyset$. \qed

\end{itemize}

\end{proof}

The following lemma is an abstraction of
 several counting arguments that we will use afterwards. 
 
\begin{lemma} Let each class $C^i_j$ be associated with
two numbers $\alpha^i_j$ and $\beta^i_j$ and $q^{-}(C^i_j) \leq \alpha^i_j, \beta^i_j \leq q^{+}(C^i_j)$. Given any non-leaf class $C^i_j$, $\alpha^i_j =  \sum_{C^i_k \in c(C^i_j)}
\alpha^i_k$ and $\beta^i_j \geq \sum_{C^i_k \in c(C^i_j)}\beta^i_k$; moreover, 
if $\beta^i_j = \sum_{C^i_k \in c(C^i_j)}\beta^i_k$ then such a non-leaf class $C^i_j$ is said to be tight in $\beta$. If $\beta^i_j> q^{-}(C^i_j)$, then $C^i_j$ has to  be tight in $\beta$.  

\begin{enumerate}

\item[(i)] Given a non-leaf class $C^i_{a^{\dagger}l^{\dagger}}$ with 
$\alpha^i_{a^{\dagger}l^{\dagger}} < \beta^i_{a^{\dagger}l^{\dagger}}$, we can find a sequence of classes $C^i_{a^{\dagger}l^{\dagger}} \supset \cdots \supset 
C^i_{a^{\dagger}1}$, where $\alpha^{i}_{a^{\dagger}t} < \beta^{i}_{a^{\dagger}t}$, for $1 \leq t \leq l^{\dagger}$.

\item[(ii)] Given a non-leaf class $C^i_x$ with $\alpha^i_x \leq \beta^i_x$, suppose that there exists a leaf class $C^i_{a^{\phi}1} \subset C^i_x$ such that $\alpha^{i}_{a^{\phi}1} > \beta^{i}_{a^{\phi}1}$. Moreover, all classes $C^i_{a^{\phi}t}$ are tight in $\beta$, where $C^i_{a^{\phi}1} \subseteq C^i_{a^{\phi}t} \subseteq C^i_x$, 
then we can find a class $C^i_{x'}$, where $C^i_{a^{\phi}1} \subset C^i_{x'} \subseteq C^i_x$, 
$\alpha^{i}_{x'} \leq \beta^i_{x'}$, and two sequences of classes with the following properties: 

\begin{enumerate}
\item[(iia)] $C^i_{a^{\phi}1} \subset C^i_{a^{\phi}2} \subset \cdots \subset C^i_{a^{\phi}l^{\phi}} \subset C^i_{x'}$, where $\alpha^i_{a^{\phi}t} >  \beta^i_{a^{\phi}t}$ for $1 \leq t \leq l^{\phi}$;

\item[(iib)] $C^i_{x'} \supset C^i_{a^{\dagger}l^{\dagger}} \supset \cdots \supset 
C^i_{a^{\dagger}1}$, where $\alpha^{i}_{a^{\dagger}t} < \beta^{i}_{a^{\dagger}t}$, for $1 \leq t \leq l^{\dagger}$. 

\end{enumerate}
\end{enumerate}

\label{lem:oftenUsedLemma}
\end{lemma}

\begin{proof} For \textrm{(i)}, since $q^{-}(C^i_{a^{\dagger}l^{\dagger}}) \leq \alpha^i_{a^{\dagger}l^{\dagger}} < \beta^i_{a^{\dagger}l^{\dagger}}$, class 
$C^i_{a^{\dagger}l^{\dagger}}$ is tight in $\beta$. Therefore, 
$\sum_{C^i_k \in c(C^i_{a^{\dagger}l^{\dagger}})}\alpha^i_k = \alpha^i_{a^{\dagger}l^{\dagger}} < \beta^i_{a^{\dagger}l^{\dagger}}
=  \sum_{C^i_k \in c(C^i_{a^{\dagger}l^{\dagger}})}\beta^i_k$. By counting, there exists a class $C^i_{a^{\dagger}(l^{\dagger}-1)} \in c(C^i_{a^{\dagger}l^{\dagger}})$ with 
$q^{-}(C^i_{a^{\dagger}(l^{\dagger}-1)}) \leq \alpha^i_{a^{\dagger}(l^{\dagger}-1)} < \beta^i_{a^{\dagger}(l^{\dagger}-1)}$. Repeating the same argument gives us the sequence of classes. 

For \textrm{(ii)}, let us climb up the tree from $C^i_{a^{\phi}1}$ until we meet a class $C^i_{x'} \subseteq C^i_x$ with $\alpha^{i}_{x'} \leq \beta^i_{x'}$. This gives us the sequence of classes stated in \textrm{(iia)}. 

Now since the class $C^i_{x'}$ is tight in $\beta$, $\sum_{C^i_k \in c(C^i_{x'})}\alpha^i_k = \alpha^i_{x'}\leq \beta^i_{x'} = \sum_{C^i_k \in c(C^i_{x'})}\beta^i_k$. 
Moreover, as $C^i_{a^{\phi}l^{\phi}} \in c(C^i_{x'})$ and $\alpha^{i}_{a^{\phi}l^{\phi}} > 
\beta^{i}_{a^{\phi}l^{\phi}}$, by counting, we can find another class $C^i_{a^{\dagger}l^{\dagger}} \in c(C^i_{x'})\backslash \{C^i_{a^{\phi}l^{\phi}}\}$ such that
$\beta^{i}_{a^{\dagger}l^{\dagger}} > \alpha^{i}_{a^{\dagger}l^{\dagger}} \geq q^{-}(C^i_{a^{\dagger}l^{\dagger}})$. 
Now applying \textrm{(i)} gives us the sequence of classes in \textrm{(iib)}. \qed
  \end{proof}

We say that $(i;a)$ is a \emph{stable pair} if there exists any stable matching in which applicant is assigned to institute $i$. A stable pair is \emph{by-passed} if institute $i$ rejects applicant $a$ during the execution of our algorithm. 

\begin{lemma} During the execution of the algorithm, if an applicant $a^{\phi}$ is rejected by institute $i$, then $(i;a^{\phi})$ is not a stable pair.

\label{lem:stablePair}
\end{lemma}

\begin{proof} We prove by contradiction. Assume that $(i;a^{\phi})$ is the first by-passed stable pair and there exists a stable matching $\mu^{\phi}$ in which $\mu^{\phi}(a^{\phi})=i$. For each class $C^i_j \in \C(i)$, we associate two numbers $\alpha^i_j:= |\mu^{\phi}(i) \cap C^i_j|$ and $\beta^i_j := |\mu(i) \cap C^i_j |  + \D(C^i_j)$. Here $\D(\cdot)$s are the values recorded in the algorithm right after $a^{\phi}$ is rejected (before the end of the while loop); similarly, $\mu(i)$ is the assignment of $i$ at that point. 

It is obvious that $\alpha^i_{a^{\phi}1} > \beta^i_{a^{\phi}1}$ and the class $C^i_{x}$ causing $a^{\phi}$ to be rejected is not $C^i_{a^{\phi}1}$. By Lemma~\ref{lem:Invariant}\textrm{(iv)}, all classes $C^i_{a^{\phi}t}$ are tight in $\beta$, where $C^i_{a^{\phi}1} \subset C^i_{a^{\phi}t} \subseteq C^i_x$. It can be checked all the conditions as stated in Lemma~\ref{lem:oftenUsedLemma}\textrm{(ii)} are satisfied. In particular, $\beta^i_x = q^{+}(C^i_x) \geq \alpha^i_x$; moreover, if $\beta^i_j > q^{-}(C^i_j)$, $C^i_j$ must be tight (by Lemma~\ref{lem:Invariant}\textrm{(iii)}). 

So, we can find two sequences of classes $\{C^i_{a^{\phi}t}\}_{t=1}^{l^{\phi}}$ and $\{C^i_{a^{\dagger}t}\}_{t=1}^{l^{\dagger}}$, where 
$C^i_{a^{\phi}l^{\phi}}$, $C^i_{a^{\dagger}l^{\dagger}} \in c(C^i_{x'})$ and 
$C^i_{x'} \subseteq C^i_x$, with the following properties: 

\begin{eqnarray*} 
q^{+}(C^i_{a^{\phi}t}) \geq |\mu^{\phi}(i) \cap C^{i}_{a^{\phi}t}| > |\mu(i) \cap C^i_{a^{\phi}t}| +  \D(C^i_{a^{\phi}t}) \geq q^{-}(C^i_{a^{\phi}t}), \forall t, 1 \leq t \leq l^{\phi}; \\
q^{-}(C^i_{a^{\dagger}t}) \leq |\mu^{\phi}(i) \cap C^{i}_{a^{\dagger}t}| < |\mu(i) \cap C^i_{a^{\dagger}t}| +  \D(C^i_{a^{\dagger}t}) \leq q^{+}(C^i_{a^{\dagger}t}), \forall t, 1 \leq t \leq l^{\dagger}. \\
\end{eqnarray*}

The second set of inequalities implies that the classes 
$\{C^i_{a^{\dagger}t}\}_{t=1}^{l^{\dagger}}$ are surplus in $\mu$. 
Thus there exists an applicant $a^{\dagger} \in (\mu(i) \backslash \mu^{\phi}(i)) \cap C^i_{a^{\dagger}1}$. Since $(i;a^{\phi})$ is the first by-passed stable pair, $i \succ_{a^{\dagger}} \mu^{\phi}(a^{\dagger})$ and since $a^{\phi}$ is rejected instead of $a^{\dagger}$, $a^{\dagger} \succ_{i} a^{\phi}$. Now observe the tuple $\mu^{\phi}(i)|^{a^{\phi}}a^{\dagger}$ is feasible due to the above two sets of strict inequalities. Thus we  have a group $(i; \mu^{\phi}(i)|^{a^{\phi}}a^{\dagger})$ to block $\mu^{\phi}$, a  contradiction. \qed

\end{proof}

\begin{lemma} At the termination of the algorithm, if there exists an institute $i \in \I$ such that 
$\D(C^i_{\sharp}) >0$, there is no stable matching in the given instance. 
\label{lem:noStableMatching}
\end{lemma}

\begin{proof} Suppose, for a contradiction, that there exists an institute 
$i$ with $\D(C^i_{\sharp}) >0$ and 
there is a stable matching $\mu^{\phi}$. Let 
$\mu$ be the assignment when the algorithm terminates. By Lemma~\ref{lem:stablePair}, if an applicant is unmatched in $\mu$, he cannot be assigned in $\mu^{\phi}$ either. So $|\mu^{\phi}| \leq |\mu|$. In the following, $\D(\cdot)$s refer to values recorded in the final outcome of the algorithm. Consider two cases. 

\begin{itemize} 

\item Suppose that $|\mu^{\phi}(i)| > |\mu(i) \cap C^i_{\sharp}|$. 
Then as $|\mu^{\phi}| \leq |\mu|$, 
we can find another institute $i' \neq i$ 
such that $|\mu^{\phi}(i')| < |\mu(i') \cap C^{i'}_{\sharp}|$. 
For each class $C^{i'}_j \in \C(i')$, let $\alpha^{i'}_j := |\mu^{\phi}(i') \cap C^{i'}_j|$ and 
$\beta^{i'}_j := |\mu(i') \cap C^{i'}_j| + \D(C^{i'}_j)$. It can be checked that the condition stated in Lemma~\ref{lem:oftenUsedLemma}\textrm{(i)} is satisfied (note that those $\beta^{i'}_j$ fulfill the condition due to Lemma~\ref{lem:Invariant}\textrm{(iii)}).  Therefore, we can find a sequence of classes $\{C^{i'}_{a^{\dagger}t}\}_{t=1}^{l^{\dagger}}$, where $C^{i'}_{a^{\dagger}l^{\dagger}} = C^{i'}_{\sharp}$, and 

$$ |\mu^{\phi}(i') \cap C^{i'}_{a^{\dagger}t}| < 
  |\mu(i') \cap C^{i'}_{a^{\dagger}t}|+ \D(C^{i'}_{a^{\dagger}t}) \leq q^{+}(C^{i'}_{a^{\dagger}t}), \forall t, 1 \leq t \leq l^{\dagger},$$ 

\noindent where the second inequality follows from Invariant~$B$. 
Then there exists an applicant $a^{\dagger} \in (\mu(i')\backslash \mu^{\phi}(i')) \cap C^{i'}_{a^{\dagger}1}$. By Lemma~\ref{lem:stablePair}, $i' \succ_{a^{\dagger}} \mu^{\phi}(a^{\dagger})$, giving us a group $(i'; \mu^{\phi}(i')|a^{\dagger})$ to block $\mu^{\phi}$, a contradiction. Note the feasibility of $\mu^{\phi}(i')|a^{\dagger}$ is due to the above set of strict inequalities.  

\item Suppose that $|\mu^{\phi}(i)| \leq |\mu(i) \cap C^{i}_{\sharp}|$. We first claim that $C^{i}_{\sharp}$ must be a surplus class in $\mu(i)$. If not, then 
$q^{-}(C^i_{\sharp}) = 
\D(C^i_{\sharp}) + |\mu(i) \cap C^i_{\sharp}| > |\mu(i) \cap C^i_{\sharp}|$, 
implying that 
$|\mu^{\phi}(i)| \geq q^{-}(C^i_{\sharp}) > |\mu(i) \cap C^{i}_{\sharp}|$, a contradiction. 
So $C^i_{\sharp}$ is a surplus class, and by Lemma~\ref{lem:Invariant}\textrm{(iii)}, 

$$|\mu^{\phi}(i)| = 
\sum_{C^i_k \in c(C^i_{\sharp})} |\mu^{\phi}(i) \cap C^i_k| \leq |\mu(i) \cap C^{i}_{\sharp}|
< |\mu(i) \cap C^{i}_{\sharp}|+ \D(C^i_{\sharp}) = \sum_{C^i_k \in c(C^i_{\sharp})} 
|\mu(i) \cap C^{i}_{k}| + \D(C^i_k). $$

For each class $C^{i}_j \in \C(i)$, let $\alpha^{i}_j := |\mu^{\phi}(i) \cap C^{i}_j|$ and $\beta^{i}_j := |\mu(i) \cap C^{i}_{j}| + \D(C^i_j)$ and invoke Lemma~\ref{lem:oftenUsedLemma}\textrm{(i)}. 
The above inequality implies that 
$\alpha^i_{\sharp} < \beta^{i}_{\sharp}$ and note that by  Lemma~\ref{lem:Invariant}\textrm{(iii)}, the condition regarding $\beta$ is satisfied. Thus we have a sequence of surplus classes $C^i_{a^{\dagger}l^{\dagger}}(=C^i_{\sharp}) \supset \cdots \supset C^i_{a^{\dagger}1}$ so that  

$$q^{-}(C^i_{a^{\dagger}t})
 \leq | \mu^{\phi}(i) \cap C^i_{a^{\phi}t}| 
< |\mu(i) \cap C^i_{a^{\dagger}t}| + \D(C^i_{a^{\dagger}t}) \leq q^{+}(C^i_{a^{\dagger}t}), \forall t, 1\leq t \leq l^{\dagger},$$

implying that there exists an applicant $a^{\dagger} \in (\mu(i) \backslash \mu^{\phi}(i)) \cap C^i_{a^{\dagger}1}$ and $i \succ_{a^{\dagger}} \mu^{\phi}(a^{\dagger})$ by virtue of Lemma~\ref{lem:stablePair}. 
The tuple $\mu^{\phi}(i)|a^{\dagger}$ is feasible because of the above set of strict inequalities. Now $(i; \mu^{\phi}(i)|a^{\phi})$ blocks $\mu^{\phi}$, a contradiction. \qed

\end{itemize}

\end{proof}

\begin{lemma} Suppose that in the final outcome $\mu$, for each institute $i \in \I$, $\D(C^i_{\sharp}) = 0$. Then $\mu$ is a stable matching. 
\label{lem:stable}
\end{lemma}

\begin{proof} For a contradiction, assume that a group $(i;g)$ blocks $\mu$. Let $a^{\phi}$ to be the highest ranking applicant in $g\backslash \mu(i)$. Since $a^{\phi}$ is part of the blocking group, he must have proposed to and been rejected by institute $i$ during the execution of the algorithm, thus $i \succ_{a^{\phi}} \mu(a^{\phi})$. By Lemma~\ref{lem:gettingBetter}, 
there exists a class 
$C^i_{a^{\phi}l^{\ddag}}$ such that 
$|\mu(i) \cap C^i_{a^{\phi}l^{\ddag}}| + \D(C^i_{a^{\phi}l^{\ddag}}) = |\mu(i) \cap C^i_{a^{\phi}l^{\ddag}}| = q^{+}(C^i_{a^{\phi}l^{\ddag}})$. Moreover, it is obvious that 
$|g \cap C^i_{a^{\phi}1}| > |\mu(i) \cap C^i_{a^{\phi}1}|$. 
We now make use of Lemma~\ref{lem:oftenUsedLemma}\textrm{(ii)} by letting $\alpha^i_j:= |g\cap C^i_j|$ and $\beta^i_j := | \mu(i) \cap C^i_j|$ for each class $C^i_j \in \C(i)$. Note that all classes are tight in $\beta$, $C^i_{a^{\phi}1} \subset C^i_{a^{\phi}l^{\ddag}}$, 
and $|\mu(i) \cap C^i_{a^{\phi}l^{\ddag}}| = 
q^{+}(C^i_{a^{\phi}l^{\ddag}}) \geq |g \cap C^i_{a^{\phi}l^{\ddag}}|$, 
satisfying all the necessary conditions. Thus, we can discover a sequence of classes 
$\{C^i_{a^{\dagger}t}\}_{t=1}^{l^{\dagger}}$
stated in Lemma~\ref{lem:oftenUsedLemma}\textrm{(iib)}, where 
$C^i_{a^{\dagger}l^{\dagger}} \in c(C^i_{a^{\phi}l})$ and  
$C^i_{a^{\phi}1} \subset C^i_{a^{\phi}l} \subseteq C^i_{a^{\phi}l^{\ddag}}$,  such that 

$$q^{-}(C^i_{a^{\dagger}t}) \leq |g \cap C^i_{a^{\dagger}t}| 
< |\mu(i) \cap C^i_{a^{\dagger}t}| \leq q^{+}(C^i_{a^{\dagger}t}), \forall j, 1\leq t \leq l^{\dagger},$$

\noindent and there exists an applicant $a^{\dagger} \in (\mu(i) \backslash g) \cap C^i_{a^{\dagger}1}$. The above set of strict inequalities mean that all classes $C^i_{a^{\dagger}t}$, $1\leq t \leq l^{\dagger}$, are surplus classes in $\mu$. 
Then $a^{\dagger}$ forms part of the affluent set $\af(C^i_{a^{\phi}l}, \mu(i))$. By Lemma~\ref{lem:gettingBetter}, 
they all rank higher than $a^{\phi}$. This contradicts our assumption that $a^{\phi}$ is the highest-ranking applicant in $g \backslash \mu(i)$. \qed

\end{proof}

\begin{lemma} Suppose that in the final outcome $\mu$, for each institute $i \in \I$, $\D(C^i_{\sharp}) = 0$. Then $\mu$ is an institute-pessimal stable matching. 
\label{lem:pessimal}
\end{lemma}

\begin{proof} Suppose, for a contradiction, 
that there exists a stable matching $\mu^{\phi}$ 
such that there exists an institute $i$ which is lexicographically better off in $\mu$ than in $\mu^{\phi}$. 
Let $a^{\dagger}$ be the highest 
ranking applicant in $\mu(i) \backslash \mu^{\phi}(i)$. 
By Lemma~\ref{lem:stablePair}, $i \succ_{a^{\dagger}} \mu^{\phi}(i)$. 
If $|\mu^{\phi}(i) \cap C^i_{a^{\dagger}t}| < 
|\mu(i) \cap C^i_{a^{\dagger}t}| 
\leq q^{+}(C^i_{a^{\dagger}t})$, for all classes $C^i_{a^{\dagger}t} \in a^{\dagger}(\C(i))$, then $(i; \mu^{\phi}(i)|a^{\phi})$ blocks $\mu^{\phi}$, a contradiction. So choose the smallest class $C^i_x \in a^{\dagger}(\C(i))$ such that $|\mu^{\phi}(i) \cap C^i_{x}| 
\geq |\mu(i) \cap C^i_x|$. It is clear that $C^i_x \supset C^i_{a^{\dagger}1}$. 

Now we apply Lemma~\ref{lem:oftenUsedLemma}\textrm{(ii)} by letting $\alpha^{i}_j := |\mu(i) \cap C^i_j|$ and $\beta^{i}_j := |\mu^{\phi}(i) \cap C^i_j|$ for each class $C^i_j \in \C(i)$. It can be checked all conditions stated in Lemma~\ref{lem:oftenUsedLemma}\textrm{(ii)} are satisfied. So there exists a class $C^i_{x'}$ such that $C^i_{a^{\dagger}1} \subset C^i_{x'} \subseteq C^i_x$ and we can find two sequences of classes $\{C^i_{a^{\phi}t}\}_{t=1}^{l^{\phi}}$ and $\{C^i_{a^{\dagger}t}\}_{t=1}^{l^{\dagger}}$, where 
$C^i_{a^{\phi}l^{\phi}}$, $C^i_{a^{\dagger}l^{\dagger}} \in c(C^i_{x'})$, with the following properties: 

\begin{eqnarray*} 
q^{+}(C^i_{a^{\dagger}t}) \geq |\mu(i) \cap C^{i}_{a^{\dagger}t}| > 
|\mu^{\phi}(i) \cap C^{i}_{a^{\dagger}t}|  \geq q^{-}(C^i_{a^{\dagger}t}), \forall t, 1 \leq t \leq l^{\dagger}; \\
q^{-}(C^i_{a^{\phi}t}) \leq |\mu(i) \cap C^{i}_{a^{\phi}t}| <  |\mu^{\phi}(i) \cap 
C^{i}_{a^{\phi}t}| \leq q^{+}(C^i_{a^{\phi}t}), \forall t, 1 \leq t \leq l^{\phi}. \\
\end{eqnarray*}

The second set of inequalities implies that we can find an applicant 
$a^{\phi} \in (\mu^{\phi}(i) \backslash \mu(i)) \cap C^i_{a^{\phi}1}$. Recall that we choose $a^{\dagger}$ to be the highest ranking applicant in $\mu(i)\backslash \mu^{\phi}(i)$, so 
$a^{\dagger} \succ_{i} a^{\phi}$. Now we have a group $(i; \mu^{\phi}(i)|^{a^{\phi}}a^{\dagger})$ to block $\mu^{\phi}$ to get a contradiction. The feasibility of 
$\mu^{\phi}(i)|^{a^{\phi}}a^{\dagger}$ is due to the above two sets of strict inequalities. \qed

\end{proof}

Based on Lemmas~\ref{lem:stablePair}, \ref{lem:noStableMatching},~\ref{lem:stable}, and~\ref{lem:pessimal}, we can draw the conclusion in this section. 

\begin{theorem} In $O(m^2)$ time, where $m$ is the total size of all preferences, the proposed algorithm discovers the applicant-optimal-institute-pessimal stable matching if stable matchings exist in the given \textbf{LCSM} instance; otherwise, it correctly reports that there is no stable matching. Moreover, if there is no lower bound on the classes, there always exists a stable matching. 
\label{thm:mainTheorem}
\end{theorem}

To see the complexity, first note that there can be only $O(m)$ proposals. The critical thing in the implementation of our algorithm is to find out the lowest ranking applicant in each affluent set efficiently. This can be done by remembering the lowest ranking applicant in each class and this information can be updated in each proposal in $O(m)$ time, since the number of classes of each institute is $O(m)$, given that the classes form a laminar family.

\subsection{Structures of Laminar Classified Stable Matching} 
\label{sec:ruralHospital}
Recall that we define the ``absorption'' operation as follows. Given a family of classes $B$, 
$\Re(B)$ returns the set of classes which are not entirely contained in other classes in $B$. 
Note that in \textbf{LCSM}, $\Re(B)$ will be composed of a pairwise disjoint set of classes. 

We review the well-known rural hospitals theorem~\cite{Gale:1985b,Roth:1986}.

\begin{theorem} (Rural Hospitals Theorem) In the {\sc hospitals/residents} 
problem, the following holds.

\begin{enumerate}[(i)]

\item A hospital gets the same number of residents in all stable matchings, and
as a result,
all stable matchings are of the same cardinality.

\item A resident who is assigned in one stable matching gets assigned
in all other stable matchings; conversely, an unassigned resident
in a stable matching remains unassigned in all other stable matchings.

\item An under-subscribed hospital gets the same set of residents in all other 
stable matchings.

\end{enumerate}
\label{thm:rural}
\end{theorem}

It turns out that rural hospitals theorem can be generalized in \textbf{LCSM}. 
On the other hand, if some institutes use intersecting classes in their 
classifications, rural hospitals theorem fails (stable matching size may differ). See the 
appendix for such an example.

\begin{theorem} (Generalized Rural Hospitals Theorem in \textbf{LCSM}) 
Let $\mu$ be a stable matching. Given any institute $i$, suppose that $B$ is 
the set of bottleneck classes in $\mu(i)$ and $D$ 
is the subset of classes in $\C(i)$ such that $\Re(B) \cup D$ partitions $\LL^i$. The following holds. 

\begin{enumerate}[(i)]

\item An institute gets the same number of applicants 
in all stable matchings, and as a result, 
all stable matchings are of the same cardinality.

\item  An applicant who is assigned in one stable matching gets assigned 
in all other stable matchings; conversely, an unassigned applicant
in a stable matching remains unassigned in all other stable matchings.

\item Every class $C^i_k \in \Re(B) \cup D$ has the same number of applicants in all stable matchings.  

\item In a class $C^i_k \subseteq C \in D$, or in a class $C^i_k$ which contains only classes in $D$, the same set of applicant in class $C^i_k$ 
will be assigned to institute $i$ in all stable matchings. 

\item A class $C^i_k$ can have different sets of applicants in different stable matchings 
only if $C^i_k \subseteq C \in \Re(B)$ or $C^i_k \supseteq C \in \Re(B)$.  

\end{enumerate}
\label{the:strongerRuralHospital}
\end{theorem}

\begin{proof} We choose $\mu^{\dagger}$ to
be the applicant-optimal stable matching.

\begin{quote} \textbf{Claim~$A$}: Suppose that $a \in \mu^{\dagger}(i) \backslash \mu(i)$.
Then there exists a class $C^i_{al} \in a(\C(i))$
such that \textrm{(i)} $|\mu(i) \cap  C^i_{al}| =q^{+}(C^i_{al})$, and \textrm{(ii)}
$a \in C^i_{al} \subseteq C \in \Re(B)$.
\end{quote}

\textbf{Proof of Claim~$A$.} If for all classes $C^i_{at} \in a(\C(i))$,
$|\mu(i) \cap C^i_{at} | < q^{+}(C^i_{at})$, then as $\mu^{\dagger}$ is applicant-optimal,
$ i \succ_{a} \mu(a)$, so $(i; \mu(i)|a)$ blocks $\mu$,
a contradiction. This establishes \textrm{(i)}.\textrm{(ii)} follows easily. \qed

Let $\tilde{B} \subseteq B$ be the subset of these bottleneck classes containing at least one applicant  $\mu^{\dagger}(i)\backslash \mu(i)$. By Claim~$A$\textrm{(ii)}, $\Re(\tilde{B})
 \subseteq \Re(B)$. This implies that
for all classes $C^i_k \in (\Re(B)\backslash \Re(\tilde{B}))\cup D$, $|\mu(i) \cap C^i_k| \geq  |\mu^{\dagger}(i) \cap C^i_k|$. Combining this fact with  Claim~$A$\textrm{(ii)}, we have

\begin{eqnarray}
|\mu(i)| & = & \sum_{C^i_k \in  (\Re(B)\backslash \Re(\tilde{B}))\cup D}|\mu(i) \cap C^i_k| +  \sum_{C^i_k \in \Re(\tilde{B})}|\mu(i) \cap C^i_k| \nonumber \\
&  \geq  & \sum_{C^i_k \in  (\Re(B)\backslash \Re(\tilde{B}))\cup D}|\mu^{\dagger}(i) \cap C^i_k| + \sum_{C^i_k \in \Re(\tilde{B})}q^{+}(C^i_{al}) \mbox{\hspace*{0.8in} (*)} \label{equ:hospital} \nonumber \\
&  =  & \sum_{C^i_k \in  (\Re(B)\backslash \Re(\tilde{B}))\cup D}|\mu^{\dagger}(i) \cap C^i_k| + \sum_{C^i_k \in \Re(\tilde{B})}|\mu^{\dagger}(i) \cap C^i_k| \nonumber \\
&= & |\mu^{\dagger}(i)|. \nonumber
\label{equ:theSameSizeInRural}
\end{eqnarray}

Thus, $|\mu| \geq |\mu^{\dagger}|$ and it cannot happen that $|\mu| > |\mu^{\dagger}|$, otherwise, there exists an applicant who is assigned in $\mu$ but not in $\mu^{\dagger}$. This contradicts the assumption that the latter is applicant-optimal. This completes the proof of  \textrm{(i)} and \textrm{(ii)} of the theorem. 

Since $|\mu|= |\mu^{\dagger}|$, Inequality~(*)  holds with equality.  We make two observations here. 

\begin{quote} \textbf{Observation}~1: For each class $C^i_k \in \Re(B)$, it is also a bottleneck in $\mu^{\dagger}(i)$. 

\textbf{Observation}~2: an applicant $a \in \mu^{\dagger}(i) \backslash \mu(i)$ must belong to a bottleneck class in $\mu^{\dagger}(i)$. 
\end{quote}

Let $B^{\dagger}$ be the set of bottleneck classes in $\mu^{\dagger}(i)$ and choose $D^{\dagger}$ so that $\Re(B^{\dagger}) \cup D^{\dagger}$ partitions $\LL^i$. By Observation~2, each applicant in $\mu^{\dagger}(i) \cap  C^i_k$, where $C^i_k \in D^{\dagger}$, must be part of $\mu(i)$. So for each class $C^i_k \in D^{\dagger}$, $|\mu(i) \cap C^i_k | \geq |\mu^{\dagger}(i) \cap C^i_k |$. We claim that it cannot happen that $|\mu(i) \cap C^i_k | > |\mu^{\dagger}(i) \cap C^i_k |$. Suppose not. Then by \textrm{(i)}, there are two possible cases. 

\begin{itemize}
\item There exists another class $C^i_{k'} \in D^{\dagger}$ so that $|\mu(i) \cap C^i_{k'} | < |\mu^{\dagger}(i) \cap C^i_{k'} |$. Then we have a contradiction to Observation~2.  

\item There exists another class $C^i_{k'} \in \Re(B^{\dagger})$ so that $|\mu(i) \cap C^i_{k'} | < |\mu^{\dagger}(i) \cap C^i_{k'} |$. For each class $C^i_j \in \C(i)$, let $\alpha^i_j := |\mu(i) \cap C^i_j|$ and $\beta^i_j := |\mu^{\dagger}(i) \cap C^i_j|$. Then we can invoke  Lemma~\ref{lem:oftenUsedLemma}\textrm{(i)} and find an applicant $a^{\phi} \in \mu^{\dagger}(i) \backslash \mu(i)$ so that for each class $C^i_{a^{\phi}t} \in a^{\phi}(\C(i))$, 
$C^i_{a^{\phi}t} \subseteq C^i_{k'}$, $|\mu(i) \cap C^i_{a^{\phi}t}| < 
|\mu^{\dagger}(i) \cap C^i_{a^{\phi}t}| \leq q^{+}(C^i_{a^{\phi}t})$. Then by Claim~$A$\textrm{(ii)} and Observation~1, there must exist another class $C^i_{k''} \in \Re(B)$ containing $a^{\phi}$ and $C^i_{k''} \supset C^i_{k'}$. By Observation~1, $C^i_{k''}$ is also a bottleneck class in $\mu^{\dagger}(i)$. This contradicts the assumption that $C^i_{k'} \in \Re(B^{\dagger})$. 
\end{itemize}

So we have that for each class $C^i_k \in D^{\dagger}$, $|\mu(i) \cap C^i_k | = |\mu^{\dagger}(i) \cap C^i_k |$. For each class $C^i_k \in B^{\dagger}$, we can use the same argument to show that $|\mu(i) \cap C^i_k | = |\mu^{\dagger}(i) \cap C^i_k |$. This gives us \textrm{(iii)} and \textrm{(iv)}. \textrm{(v)} is a consequence of \textrm{(iv)}. \qed

\end{proof}

\section{NP-completeness of $\mathbb{P}$-Classified Stable Matching}

\label{sec:np}

\begin{theorem} Suppose that the set of posets $\mathbb{P}=\{P_1,P_2,\cdots, P_k\}$ 
contains a poset which is not a downward forest. Then it is NP-complete to decide the existence of a stable matching in $\mathbb{P}$-classified stable matching. This NP-completeness holds even if there is no lower bound on the classes. 
\label{thm:notDownwardForstIsNPC}
\end{theorem}

Our reduction is from {\sc one-in-three sat}. It is involved and technical, so we just highlight the idea here. As $\mathbb{P}$ must contain a poset that has a ``$V$'' in it, some institutes use intersecting classes. In this case, even if there is no lower bound on the classes, it is possible that the given instance disallows any stable matching. We make use of this fact to design a special gadget. The main technical difficulty of our reduction lies in that in the most strict case, we can use at most two classes in each institute's classification. 

\section{Polyhedral Approach} 
\label{sec:polyhedral}
In this section, we take a polyhedral approach to studying \textbf{LCSM}. 
We make the simplifying assumption that there is no lower bound. In this scenario, we can use a simpler definition to define a stable matching. 

\begin{lemma} In \textbf{LCSM}, if there is no lower bound, i.e., given any class $C^i_j$, $q^{-}(C^i_j)=0$, then a stable matching as defined in Definition~\ref{def:first} can be equivalently defined as follows. A feasible matching $\mu$ is stable if and only if there is no blocking pair. A pair $(i,a)$ is blocking, given that $\mu(i) = (a_{i1}, a_{i2},\cdots, a_{ik})$, $k \leq Q(i)$, if 

\begin{itemize}
\item $i \succ_{a} \mu(a)$; 
\item for any class $C^i_{at} \in a(\C(i))$, $|\LL^i_{\succ a} \cap \mu(i) \cap C^i_{at}| < q^{+}(C^i_{at})$. 
\end{itemize}
\label{pro:blockingPair}
\end{lemma}

The definition of blocking pairs suggests a generalization of the \emph{comb} used by Ba\"{i}ou and Balinski~\cite{Baiou:2000}. 

\begin{definition} Let $\Gamma = \I \times \A$ denote the set of acceptable institute-applicant pairs. The \emph{shaft} $S(A^i)$, based on a feasible tuple $A^i$ of institute $i$, is defined as follows:

$$S(A^i)= \{(i,a') \in \Gamma: \forall C^i_j \in a'(\C(i)), |\LL^i_{\succ a'}\cap A^i \cap C^i_{j}| < q^{+}(C^i_{j})\}.$$

The \emph{tooth} $T(i,a)$ is defined for every $(i,a) \in \Gamma$ as follows: 

$$T(i,a) = \{(i',a) \in \Gamma: i' \succeq_{a} i\}.$$ 
\label{def:comb}
\end{definition}

In words, $(i,a')$ forms part of the shaft $S(A^i)$, only if the collection of $a'$ and all applicants in $A^i$ ranking strictly higher than $a'$ does not violate the quota of any class in $a'(\C(i))$. We often refer to an applicant $a \in A^i$ as a \emph{tooth-applicant}. 

We associate a $|\Gamma|$-vector $x^{\mu}$ (or simply $x$ when the context is clear) with a matching $\mu$: $x^{\mu}_{ia}=1$ if $\mu(a)=i$, otherwise, $x^{\mu}_{ia}=0$. Suppose that $\hat{\Gamma} \subseteq \Gamma$. Then $x(\hat{\Gamma}) = \sum_{(i,a)\in \hat{\Gamma}}x_{ia}$. 
We define a \emph{comb} $K(i, S(A^i))$ as the union of the teeth $\{T(i, a_i)\}_{a_i \in A^i}$ and the shaft $S(A^i)$. 

\begin{lemma} 

Every stable matching solution $x$ satisfies the \emph{comb inequality} for any comb $K(i,S(A^i))$:
\begin{equation*}
x(K(i,S(A^i)) \equiv x(S(A^i)) + \sum_{a_j \in A^i} x(T(i,a_j)\backslash \{i, a_j\}) \geq |A^i|. \label{equ:stableConstraint}
\end{equation*}
\label{pro:stableConstraint}
\end{lemma}

It takes a somehow involved counting argument to prove this lemma. Here is the  intuition about why the comb inequality captures the stability condition of a matching. The value of the tooth $x(T(i,a))$ reflects the ``happiness'' of the applicant $a \in A^i$. If $x(T(i,a))=0$, applicant $a$ has reason to shift to institute $i$; on the other hand, the values collected from the shaft $x(S(A^i))$ indicates the ``happiness'' of institute $i$: whether it is getting enough high ranking applicants (of the ``right'' class). An overall small comb value $x(K(i,S(A^i)))$ thus expresses the likelihood of a blocking group including $i$ and some of the applicants in $A^i$. 

Now let $\mathbb{K}^i$ denote the set all combs of institute $i$. We write down the linear program:

\begin{eqnarray}
\sum_{i: (i,a) \in \Gamma} x_{ia} &\leq & 1, \forall a \in \A \label{equ:lpFirst} \\
\sum_{a: (i,a) \in \Gamma, a \in C^i_j} x_{ia} &\leq & q^{+}(C^i_j), \forall i \in \I, \forall C^i_j \in \C(i)  \label{equ:lpSecond}\\
x(K(i,S(A^i))) = \sum_{(i,a) \in K(i,S(A^i))} x_{ia} &\geq& |A^i|, \forall K(i,S(A^i)) \in \mathbb{K}^i, \forall i \in \I \label{equ:lpThird} \\
x_{ia} &\geq & 0, \forall (i,a) \in \Gamma \label{equ:lpFourth}
\end{eqnarray}

Suppose there is no classification, i.e., {\sc Hospitals/Residents} problem. Then this LP reduces to the one formulated by Ba\"{i}ou and Balinski~\cite{Baiou:2000}. However, it turns out that this polytope is \emph{not} integral. The example in Figure~\ref{fig:counterExample} demonstrates
the non-integrality of the polytope. In particular, observe that since $\mu$ is applicant-optimal, in all other stable matchings, applicant $a_3$ can only be matched to $i_5$. However, the value $x_{i_1 a_3}= 0.2>0$ indicates that $x$ is \emph{outside} of the convex hull of integral stable matchings.

\begin{figure*}[h]\footnotesize
\hrule
\begin{tabbing}
\hspace{0pt}\=\hspace{150pt}\=\hspace{150pt}\=
\hspace{70pt}\=\hspace{30pt}\=\hspace{15pt}\=\hspace{15pt}\=\hspace{15pt}\=
\hspace{15pt}\\

\> Institute Preferences \>  Classifications     \> Class bounds   \\
\> $i_1\mbox{:} a_1 a_6 a_7 a_2 a_3   $  \> $C^1_1=\{a_2,a_3\}$ \> $Q(i_1)=2$, $q^{+}(C^1_1)=1$    \\
\> $i_2\mbox{:} a_4 a_7$ \> \> $Q(i_2)=1$ \\
\> $i_3\mbox{:} a_2 a_4$ \> \> $Q(i_3)=1$ \\
\> $i_4\mbox{:} a_5 a_6$ \> \> $Q(i_4)=1$ \\
\> $i_5\mbox{:} a_3 a_5 a_7 a_1  $  \> $C^5_1=\{a_3,a_5\}$ \> $Q(i_5)=2$, $q^{+}(C^5_1)=1$    \\\\\\

\> Applicant Preferences \\
\> $a_1\mbox{:} i_5 i_1  $ \\
\> $a_2\mbox{:} i_1 i_3  $   \\
\> $a_3\mbox{:} i_1 i_5$   \\
\> $a_4\mbox{:} i_3 i_2$ \\
\> $a_5\mbox{:} i_5 i_4$ \\
\> $a_6\mbox{:} i_4 i_1$ \\
\> $a_7\mbox{:} i_2 i_1 i_5$ \\\\

\> (Applicant-optimal stable matching) $\mu = \{(i_1; a_2, a_6), (i_2;a_7),(i_3;a_4),(i_4;a_5), (i_5;a_1, a_3) \}$
 \\\\

\> (A fractional matching $x$ that is \emph{not} inside the convex hull of integral stable matchings)\\
\> $x_{i_1 a_1} =0.5$, $x_{i_1 a_6} =0.8$, $x_{i_1 a_7} =0.2$, $x_{i_1 a_2} =0.3$, $x_{i_1 a_3} =0.2$.\\
\> $x_{i_2 a_4} =0.7$, $x_{i_2 a_7} =0.3$.\\
\> $x_{i_3 a_2} =0.7$, $x_{i_3 a_4} =0.3$.\\
\> $x_{i_4 a_5} =0.8$, $x_{i_4 a_6} =0.2$.\\
\> $x_{i_5 a_3} =0.8$, $x_{i_5 a_5} =0.2$, $x_{i_5 a_7} =0.5$, $x_{i_5 a_1} =0.5$. \\

\end{tabbing}
\vspace*{-0.35in}
\caption{An example showing that the polytope formed by Constraints~(\ref{equ:lpFirst})-(\ref{equ:lpFourth}) is not integral. Since $\mu$ is applicant-optimal, in all other stable matchings, applicant $a_3$ can only be matched to $i_5$. However, the value $x_{i_1 a_3}= 0.2>0$ indicates that $x$ is \emph{outside} of the convex hull of integral stable matchings.}
\vspace*{0.03in}
\hrule
\label{fig:counterExample}
\end{figure*}

Here we make a critical observation. Suppose that in a certain matching $\mu^{\phi}$, applicant $a_3$ is assigned to $i_1$. Then $a_2$ cannot be assigned to $i_1$ due to the bound $q^{+}(C^1_1)$ (see Constraint~(\ref{equ:lpSecond})). If $\mu^{\phi}$ is to be stable, then $a_2$ must be assigned to some institute ranking higher than $i_1$ on his list (in this example there is none), 
otherwise, $(i, \mu^{\phi}(i_1)|^{a_3}a_2)$ is bound to be a blocking group in $\mu^{\phi}$. Thus, the required constraint to avoid this particular counter-example can be written as 

$$x(T(i_1, a_2)\backslash \{i_1,a_2\}) \geq x_{i_1 a_3}.$$ 

We now formalize the above observation. Given any class $C^i_j \in \C(i)$, we define a \emph{class-tuple} $t^i_j = (a_{i1}, a_{i2},\cdots, a_{iq^{+}(C^i_j)})$. Such a tuple fulfills the following two conditions: 

\begin{enumerate}
\item $t^i_j \subseteq C^i_j$; 
\item if $C^i_j$ is a non-leaf class, then given any subclass $C^i_k$ of $C^i_j$, 
$|t^i_j \cap C^i_k| \leq q^{+}(C^i_k)$. 
\end{enumerate}

Let $\LL^i_{\prec t^i_j}$ denote the set of applicants ranking lower than all applicants in $t^i_j$ and $\LL^i_{\succeq t^i_j}$ the set of applicants ranking at least as high as the lowest ranking applicant in $t^i_j$. 

\begin{lemma} 
Every stable matching solution $x$ satisfies the following inequality for any class-tuple $t^i_j$: 
\begin{equation*}
\sum_{a_{ij} \in t^i_j} x(T(i,a_{ij}) \backslash \{i,a_{ij}\}) \geq \sum_{a \in C^i_j \cap \LL^i_{\prec t^i_j}}x_{ia}. \end{equation*}
\label{pro:newConstraint}
\end{lemma}

As before, it takes a somehow involved counting argument to prove the lemma but its basic idea is already portrayed in the above example. Now let $\mathbb{T}^i_j$ denote the set of class-tuples in class $C^i_j \in \C(i)$ and  $\LL^i_{\prec t^i_j}$ denote the set of applicants ranking lower than all applicants in $t^i_j$. We add the following sets of constraints.

\begin{equation}
\sum_{a_{ij} \in t^i_j} x(T(i,a_{ij}) \backslash \{i,a_{ij}\}) \geq \sum_{a \in C^i_j \cap \LL^i_{\prec t^i_j}}x_{ia}, \forall t^i_j \in \mathbb{T}^{i}_j, \forall \mathbb{T}^i_j \label{equ:lpFifth}
\end{equation}

Let $P_{fsm}$ denote the set of all solutions satisfying~(\ref{equ:lpFirst})-(\ref{equ:lpFifth}) 
and $P_{sm}$ the convex hull of all (integral) stable matchings. 
In this section, our main result is $P_{fsm}=P_{sm}$. We say $(i,a)$ are \emph{matched under $x$} if $x_{ia}>0$. 

\begin{definition} Let $x \in P_{fsm}$ and $\Omega_i(x)$ be the set of applicants that are matched to institute $i$ under $x$. Let $\Omega_i(x)$ be composed of $a_{i1},a_{i2},\cdots$, ordered based on the decreasing preference of institute $i$.  

\begin{enumerate}
\item Define $H_{i}(x)$ as a tuple composed of applicants chosen based on the following procedure: adding $a_{ij}$ greedily unless adding the next applicant into $H_i(x)$ will cause $H_{i}(x)$ to violate the quota of some class. Equivalently, $a_{il} \not \in H_i(x)$ only if there exists a class $C^i_j \in a_{il}(\C(i))$ such that $|H_{i}(x) \cap \{a_{it}\}_{t=1}^{l-1}| = q^{+}(C^i_j)$. 

\item Define $E_i(x)$ as a tuple composed of applicants for whom institute $i$ is the most preferred institute that they are matched under $x$, i.e., an applicant $a \in E_i(x)$, if 
$x(T(i,a)\backslash \{(i,a)\})=0$ and $x_{ia}>0$. The order of the applicants in $E_i(x)$ is based on the decreasing preference of institute $i$. 

\end{enumerate} 
\label{def:EH}
\end{definition}

\begin{lemma} $E_i(x)$ is feasible for institute $i$. 
\label{lem:eix}
\end{lemma}
\begin{proof} We need to show that given any class $C^i_j \in \C(i)$, $|E_i(x) \cap C^i_j| \leq q^{+}(C^i_j)$. We proceed by induction on the height of $C^i_j$ in the tree structure of $\C(i)$. The base case is a leaf class. If $|E_i(x) \cap C^i_j| > q^{+}(C^i_j)$, form a class-tuple by picking the first $q^{+}(C^i_j)$ applicants in $E_i(x) \cap C^i_j$. Then Constraint~(\ref{equ:lpFifth}) is violated in such a class-tuple. For the induction step, if $|E_i(x) \cap C^i_j| > q^{+}(C^i_j)$, again choose the $q^{+}(C^i_j)$ highest-ranking applicants in $E_i(x) \cap C^i_j$ and we claim they form a class-tuple of $C^i_j$, the reason being that by induction hypothesis, given any $C^i_k \subset C^i_j$, $|E_i(x) \cap C^i_k| \leq q^{+}(C^i_k)$. Now Constraint~(\ref{equ:lpFifth}) is again violated in such a class-tuple. \qed

\end{proof}

\begin{lemma} Suppose that $x \in P_{fsm}$. 

\begin{enumerate}
\item[(i)] For each institute $i \in \I$, we can find two sets $U$ and $V$ of pairwise disjoint classes so that $U \cup V$ partitions $\LL^i$ and all applicants in $\Omega_i(x) \backslash H_i(x)$ belong to the classes in $U$. Moreover,

\begin{enumerate}
\item[(ia)]  $|H_i(x)| = \sum_{C^i_k \in U}q^{+}(C^i_k) + \sum_{C^i_k \in V}|H_i(x) \cap C^i_k|$; 

\item[(ib)] for each class $C^i_k \in U$, $|H_i(x) \cap C^i_k| = |E_i(x) \cap C^i_k| = q^{+}(C^i_k)$;  for each class $C^i_k \in V$ and each applicant $a \in C^i_k$, if $x_{ia}>0$, then $x_{ia}=1$; 

\item[(ic)] for each class $C^i_k \in U$, $\sum_{a \in C^i_k}x_{ia}= q^{+}(C^i_k)$. 

\end{enumerate}

\item[(ii)] For every applicant $a \in H_{i}(x)$, $x(T(i,a))= \sum_{i \in \I}x_{ia}=1$; moreover, given any two institutes $i$, $i' \in \I$, $H_i(x) \cap H_{i'}(x) = \emptyset$. 

\item[(iii)] $|H_i(x)| = |E_i(x)|$ for all institutes $i \in \I$. 

\item[(iv)] $\sum_{a \in \A}x_{ia} = |E_i(x)|$ for all institutes $i \in \I$. 
\end{enumerate}

\label{lem:fractuals}
\end{lemma}

\begin{proof} For \textrm{(i)}, given any applicant $a \in \Omega_i(x) \backslash H_i(x)$, by Definition~\ref{def:EH}, there exists some class $C^i_j \in a(\C(i))$ for which $|H_i(x) \cap C^i_j| = q^{+}(C^i_j)$. Let $B$ be the set of classes $C^i_j$ which contain at least one applicant in $\Omega_i(x) \backslash H_i(x)$ and $|C^i_j \cap H_i(x)|=q^{+}(C^i_j)$. Let $U:= \Re(B)$ and choose $V$ in such a way so that $U \cup V$ partitions $\LL^i$. Now \textrm{(ia)} is a consequence of counting. We will prove \textrm{(ib)(ic)} afterwards. 

For \textrm{(ii)}, by definition of $H_i(x)$, none of the applicants in $\Omega_i(x) \backslash H_i(x)$ contributes to the shaft $x(S(H_i(x)))$. As a result, for Constraint~(\ref{equ:lpThird}) to hold for the comb $K(i,S(H_i(x)))$, every tooth-applicant $a \in H_i(x)$ must contribute at least 1, and indeed, by Constraint~(\ref{equ:lpFirst}), exactly 1. So we have the first statement of \textrm{(ii)}. The second statement holds because it cannot happen that $x(T(i,a)) = x(T(i',a))=1$, given that $x_{ia}>0$ and $x_{i'a}>0$. 

For \textrm{(iii)}, By Definition~\ref{def:EH}, all sets $E_i(x)$ are disjoint; thus, every applicant who is matched under $x$ belongs to exactly one $E_i(x)$ and at most one $H_i(x)$ by \textrm{(ii)}. Therefore, $\sum_{i \in \I}|E_i(x)| \geq 
\sum_{i \in \I}|H_i(x)|$ and we just need to show that for each institute $i$, $|E_i(x)| \leq |H_i(x)|$, and this follows by using \textrm{(ia)}:

\begin{equation}
|H_i(x)| = \sum_{C^i_k \in U}q^{+}(C^i_k) +  \sum_{C^i_k \in V} |H_i(x) \cap C^i_k|  \geq 
 \sum_{C^i_k \in U}|E_i(x) \cap C^i_k|+ \sum_{C^i_k \in V}|E_i(x) \cap C^i_k|    = |E_i(x)|, 
\label{equ:here}
\end{equation}

\noindent where the inequality follows from Lemma~\ref{lem:eix} and the fact all applicants in $\Omega_i(x) \backslash H_i(x)$ are in classes in $U$. So this establishes \textrm{(iii)}. Moreover, as Inequality~(\ref{equ:here}) must hold with equality throughout, for each class $C^i_k \in V$, if applicant $a \in C^i_k$ is matched to institute $i$ under $x$, he must belong to both $H_i(x)$ and $E_i(x)$, implying $x_{ia}=1$; given any class $C^i_k \in U$, $|H_i(x) \cap C^i_k|= 
|E_i(x) \cap C^i_k| = q^{+}(C^i_k)$. So we have \textrm{(ib)}. 

For \textrm{(iv)}, consider the comb $K(i,S(E_i(x)))$. By definition, $x(T(i,a)\backslash \{(i,a)\})=0$ for each applicant $a \in E_i(x)$. So 

\begin{eqnarray*}x(K(i,S(E_i(x)))) &=& x(S(E_i(x))) \\
&=& \sum_{C^i_k \in V}|E_i(x) \cap C^i_k| +  \sum_{C^i_k \in U} 
\sum_{a' \in C^i_k, (i,a') \in S(E_i(x))}x_{ia'} \\
&\leq&  \sum_{C^i_k \in V}|E_i(x) \cap C^i_k| + 
\sum_{C^i_k \in U}q^{+}(C^i_k) = |E_i(x)|,
\end{eqnarray*}

\noindent 
where the inequality follows from Constraint~(\ref{equ:lpSecond}) and the rest can be deduced from \textrm{(ib)}. By Constraint~(\ref{equ:lpThird}), the above inequality must hold with equality. So for each class $C^i_k \in U$, 
$\sum_{a' \in C^i_k, (i,a') \in S(E_i(x))}x_{ia'} = \sum_{a' \in C^i_k}x_{ia'}= q^{+}(C^i_k)$, giving us \textrm{(ic)} and implying that there is no applicant in $C^i_k \in U$ who is matched to institute $i$ under $x$ ranking lower than all applicants in $E_i(x) \cap C^i_k$. The proof of \textrm{(iv)} follows by 

$$ \sum_{a \in \A}x_{ia}= \sum_{C^i_k \in V}\sum_{a \in C^i_k}x_{ia} + 
\sum_{C^i_k \in U}\sum_{a \in C^i_k}x_{ia} = \sum_{C^i_k \in V}|E_i(x) \cap C^i_k|+ \sum_{C^i_k \in U}q^{+}(C^i_k) = |E_i(x)|.$$ \qed
\end{proof}

\subsection*{Packing Algorithm}

We now introduce a packing algorithm
 to establish the integrality of the polytope. Our algorithm is generalized from that proposed by Sethuraman, Teo, and Qian~\cite{Sethuraman:2006a}. Given $x \in P_{fsm}$, for each institute $i$, we create $|E_i(x)|$ ``bins,'' each of size (height) 1; each bin is indexed by $(i,j)$, where $1 \leq j \leq |E_i(x)|$. Each $x_{ia}>0$ is an ``item'' to be packed into the bins. Bins are filled from the bottom to the top. When the context is clear, we often refer to those items $x_{ia}$ as simply applicants; if applicant $a \in C^i_j$, then the item $x_{ia}$ is said to belong to the class $C^i_j$. 

In Phase~0, each institute $i$ puts the items $x_{ia}$, if $a \in H_i(x)$, into each of its $|E_i(x)|$ bins. In the following phase, $t=1,2,\cdots$, our algorithm proceeds by 

\begin{itemize}
\item first finding out the set $L_t$ of bins with maximum available space;
\item then assigning each of the bins in $L_t$ one item.

\end{itemize}

The assignment in each phase proceeds by \emph{steps}, indexed by $l=1,2,\cdots, |L_t|$. The order of the bins in $L_t$ to be examined  does not matter. 
How the institute $i$ chooses the items to be put into its bins is the crucial part in which our algorithm differs from that of Sethuraman, Teo, and Qian. 
We maintain the following invariant.

\begin{quote} \textbf{Invariant}~$C$: The collection of the least preferred items in the $|E_i(x)|$ bins (e.g., the items currently on top of institute $i$'s bins) should respect of the quotas of the classes in $\C(i)$. 
\end{quote}

Subject to this invariant, institute $i$ chooses the best remaining item and adds it into the bin $(i,j)$, which has the maximum available space in the current phase. This unavoidably raises another issue: how can we be sure that there is at least one remaining item for institute $i$ to put into the bin $(i,j)$ without violating Invariant~$C$? We will address this issue in our proof.


\begin{theorem} Let $x \in P_{fsm}$. Let $M_{i,j}$ be the set of applicants assigned to bin $(i,j)$ at the end of any step of the packing procedure and $a_{i,j}$ be the lowest-ranking applicant of institute $i$ in bin $(i,j)$ (implying $x_{ia_{i,j}}$ is on top of bin $(i,j)$). Then 

\begin{enumerate}

\item[(i)] In any step, suppose that the algorithm is examining bin $(i,j)$. Then institute $i$ can find at least one item in its remaining items to add into bin $(i,j)$ without violating Invariant~$C$;

\item[(ii)] For all bins $(i,j)$, $x(M_{i,j}\backslash \{a_{i,j}\}) + x(T(i, a_{i,j})) = 
x(M_{i,j}) + x(T(i, a_{i,j})\backslash \{(i, a_{i,j}) \}) =1$; 

\item[(iii)] At the end of any step, institute $i$ can organize a comb $K(i, S(A^i))$ where 
$A^i$ is composed of applicants in $\{a_{i,j'}\}_{j'=1}^{|E_{i}(x)|}$ so that 
$x(K(i, S(A^i))= \sum_{j'=1}^{|E_i(x)|}x(M_{i,j'}) + \sum_{j'=1}^{|E_i(x)|}x(T(i,a_{i,j'})\backslash \{(i,a_{i,j'})\})= |E_i(x)|$;

\item[(iv)] At the end of any step, an item $x_{ia}$ is not put into institute $i$'s bins if and only if there exists a class $C^i_{at} \in a(\C(i))$ so that $|\{a_{i,j'}\}_{j'=1}^{|E_i(x)|} \cap C^i_{at} \cap \LL^i_{\succ a}| = q^{+}(C^i_{at})$. 

\item[(v)] If $x_{ia}$ is packed and $x_{i'a}$ is not, then $i' \succ_{a} i$;

\item[(vi)] At the end of any phase, the $a_{i,j}$ in all bins are distinct. In particular, for any applicant $a$ who is matched under $x$, there exists some bin $(i,j)$ such that $a = a_{i,j}$. 

\end{enumerate}

\label{thm:packing} 
\end{theorem} 

\begin{proof} We first assume that \textrm{(ii)} holds and prove \textrm{(i)}. Observe that \textrm{(ii)} implies that given any applicant $a \in E_i(x)$, its corresponding item $x_{ia}$, if already put into a bin, must be on its top and fills it completely. Since $(i,j)$ currently has available space, at least one applicant in $E_i(x)$ is not in institute $i$'s bins yet. We claim that there exists at least one remaining applicant in $E_i(x)$ that can be added into bin $(i,j)$. Suppose not. Let the set of applicants in $E_i(x)$ that are not put into $i$'s bins be $G$. 
Given any applicant $a \in G$, there must exist some class $C^i_k \in a(\C(i))$ for which $|\bigcup_{1 \leq j' \leq |E_i(x)|, j'\neq j}a_{i,j'} \cap C^i_k| = q^{+}(C^i_k)$. Let $B$ be the set of classes $C^i_k$ that contains at least one applicant in $G$ and $|\bigcup_{1 \leq j' \leq |E_i(x)|, j'\neq j}a_{i,j'} \cap C^i_k| = q^{+}(C^i_k)$. Let $G'$ be 
$(E_i(x)  \backslash G)\backslash \bigcup_{C^i_k \in \Re(B)}C^i_k$, the subset of applicants in $E_i(x)$ that are already put into the bins but not belonging to any class in $\Re(B)$. Note that none of the applicants in $G'$ can be in the bin $(i,j)$. Thus, by counting the number of the bins minus $(i,j)$, we have 

$$ |E_i(x)| - 1 \geq  |G'| + \sum_{C^i_k \in \Re(B)}|\bigcup_{j'=1, j'\neq j}^{|E_i(x)|}a_{i,j'} \cap C^i_k| = 
|G'| + \sum_{C^i_k \in \Re(B)}q^{+}(C^i_k)  $$ 

Note that all applicants in $E_{i}(x) \backslash {G'}$ are in some class in $\Re(B)$ (either they are already put into the bins or not). Then by the pigeonhole principle, there is at least one class $C^i_k \in \Re(B)$ for which $|(E_{i}(x) \backslash G') \cap C^i_k| > q^{+}(C^i_k)$, contradicting Lemma~\ref{lem:eix}. 

We now prove \textrm{(ii)-(vi)} by induction on the number of phases. In the beginning, \textrm{(ii)(v)(vi)} holds by Lemma~\ref{lem:fractuals}{(ii)(iii)}. \textrm{(iii)(iv)} hold by 
setting $A^i := H_i(x)$ and observation Definition~\ref{def:EH} and Lemma~\ref{lem:fractuals}\textrm{(ii)}. Suppose that the theorem holds up to Phase~$t$. Let $\alpha$ be the maximum available space in Phase~$t+1$. Suppose that the algorithm is examining bin $(i,j)$ and institute $i$ chooses item $x_{ia}$ to be put into this bin. From \textrm{(vi)} of the induction hypothesis, applicant $a$ is on top of another bin $(i',j')$, where $i' \neq i$, in the beginning of phase $t+1$. Then by \textrm{(ii)(v)} of the induction hypothesis, 

\begin{equation}
x(T(i,a)) \leq x(T(i',a)) - x_{i'a} = 1 - x(M_{i',j'}) \leq \alpha, 
\label{equ:simple1}
\end{equation}

\noindent where the last inequality follows from our assumption that in Phase~$t+1$, the maximum available space is $\alpha$. Note also that 

\begin{equation} x(T(i,a)) = \alpha, \mbox{ then } (i',j') \in L_{t+1} \mbox{ (bin $(i',j')$ is also examined in Phase $t+1$). } 
\label{equ:simple2}
\end{equation}

Assume that $\overline{A}^{i}$ is a tuple composed of applicants in 
$\{a_{i,j'}\}_{j'=1}^{|E_i(x)|}$. For our induction step, let $A^i := \overline{A}^{i}|^{a_{i,j}}
a$, the new set of items on top of $i$'s bins after $a$ is put on top of $a_{i,j}$. 

We first prove \textrm{(iv)}. Since $x_{ia}$ is not put into the bin before this step, by \textrm{(iv)} of the induction hypothesis, there exists some class $C^i_{al} \in a(\C(i))$ for which $|\overline{A}^i \cap C^i_{al} \cap \LL^i_{\succ a}|=q^{+}(C^i_{al})$. Let $C^i_{al}$ be the smallest such class. Since $x_{ia}$ is allowed to put on top of $x_{ia_{i,j}}$, 
$a_{ij} \succ_{i} a$ and $a_{ij} \in C^i_{al}$, otherwise, Invariant~$C$ regarding $q^{+}(C^i_{al})$ is violated. 

Now we show that all other items $x_{ia'}$ fulfill the condition stated in \textrm{(iv)}. There are two cases.

\begin{itemize}

\item Suppose that $x_{ia'}$ is not put into the bins yet.

\begin{itemize}
\item Suppose that $a_{i,j} \succ_{i} a' \succ_{i} a$. We claim that it cannot happen that 
for all classes $C^i_{a't} \in a'(\C(i))$, $|A^i \cap C^i_{a't} \cap \LL^i_{\succ a'}| < q^{+}(C^i_{a't})$, otherwise, $A^i|^{a}a'$ is still feasible, in which case institute $i$ would have chosen $x_{ia'}$, instead of $x_{ia}$ to put into bin $(i,j)$, a contradiction. 

\item Suppose that $a_{i,j} \succ_{i} a \succ_{i} a'$. By \textrm{(iv)} of the induction hypothesis, 
there exists a class $C^i_{a'l'} \in a'(\C(i))$ for which 
$|\overline{A}^i \cap C^i_{a'l'} \cap \LL^i_{\succ a'}| = q^{+}(C^i_{a'l'})$. If 
$C^i_{a'l'} \not \subset C^i_{al}$, it is easy to see that 
$|A^i \cap C^i_{a'l'} \cap \LL^i_{\succ a'}| = q^{+}(C^i_{a'l'})$; 
if $C^i_{a'l'} \subset C^i_{al}$, then 
$C^i_{al} \in a'(\C(i))$ and we have $|A^i \cap C^i_{al} \cap \LL^i_{\succ a'}| = q^{+}(C^i_{al})$. In both situations, the condition of \textrm{(iv)} regarding $x_{ia'}$ is satisfied. 

\end{itemize}

\item Suppose that $x_{ia'}$ is already put into the bins. It is trivial if $a' \succ_{i} a$, so assume that $a \succ_{i} a'$. We claim that none of the classes $C^i_{a't} \in a'(\C(i))$ can be a subclass of $C^i_{al}$ or $C^i_{al}$ itself. Otherwise, $C^i_{al} \in a'(C(i))$, and we have $q^{+}(C^i_{al})= |\overline{A}^i \cap C^i_{al} \cap \LL^i_{\succ a}| \geq |\overline{A}^i \cap C^i_{al} \cap \LL^i_{\succ a'}|$, a contradiction to \textrm{(iv)} of the induction hypothesis. Now since for every class $C^i_{a't} \in a'(\C(i))$, we have $C^i_{a't} \not \subseteq C^i_{al}$, we have $|A^i \cap C^i_{a't} \cap \LL^i_{\succ a'}| = 
|\overline{A}^i \cap C^i_{a't} \cap \LL^i_{\succ a'}| < q^{+}(C^i_{a't})$, where the strict inequality is due to the induction hypothesis. 

\end{itemize}

We notice that the quantity $\sum_{j'=1}^{|E_i(x)|}x(M_{i,j'})$ is exactly the sum of the shaft 
$x(S(\overline{A}^i))$ (before $x_{ia}$ is added) or $x(S(A^i))$ (after $x_{ia}$ is added) 
by observing \textrm{(iv)}. Below let $x(\overline{M}_{i,j})$ and $x(M_{i,j})$ denote the total size of the items in bin $(i,j)$ before and after $x_{ia}$ is added into it. So 
$x(M_{i,j}) = x(\overline{M}_{i,j}) + x_{ia}$. Now we can derive the following: 

\begin{eqnarray*}
x(K(i,S(A^i)))
 & = & x(S(A^i)) + x(T(i,a)\backslash \{(i,a)\}) + \sum_{j'=1, j'\neq j}^{|E_i(x)|}
x(T(i,a_{i,j'})\backslash \{(i,a_{i,j'})\})  \\
& = & x(\overline{M}_{i,j}) + x_{ia} +  x(T(i,a) \backslash \{(i,a)\}) + \sum_{j'=1, j'\neq j}^{|E_i(x)|}x(M_{i,j'}) + x(T(i,a_{i,j'})\backslash \{(i,a_{i,j'})\})  \\
& = & x(\overline{M}_{i,j}) + x(T(i,a)) + |E_i(x)|-1  \mbox{ \hspace*{0.1in}(by \textrm{(ii)} of the induction hypothesis)} \\
& \geq & |E_i(x)| \mbox{ \hspace*{0.1in}(by Constraint~(\ref{equ:lpThird}))}
\end{eqnarray*}

For the above inequality to hold,  

\begin{equation}
x(\overline{M}_{i,j}) + x(T(i,a)) \geq 1. 
\label{equ:simple3}
\end{equation}

Since $x(\overline{M}_{i,j}) = 1 - \alpha$ and $x(T(i,a)) \leq \alpha$ by Inequality~(\ref{equ:simple1}), Inequality~(\ref{equ:simple3}) must hold with equality, implying that $x(K(i,S(A^i))) = |E_i(x)|$, giving us \textrm{(iii)}.  

Since institute $i$ puts $x_{ia}$ into bin $(i,j)$, the ``new'' $M_{i,j}$ and the ``new'' $a_{i,j}$ (=$a$) satisfies 

$$x(M_{i,j}) + x(T(i,a)\backslash \{(i,a)\}) = 1.$$ 

This establishes \textrm{(ii)}. \textrm{(v)} follows because 
Inequality~(\ref{equ:simple1}) must hold with equality throughout. Therefore, there is no institute $i''$ which ranks strictly between $i$ and $i'$ and $x_{i''a}>0$. 

Finally for \textrm{(vi)}, note that $x(T(i,a))= \alpha$ if the item $x_{ia}$ is put into some bin in Phase~$t+1$. All such items are the least preferred items in their respective ``old'' bins (immediately before Phase~$t+1$), it means the items on top of the newly-packed bins are still distinct. Moreover, from~(\ref{equ:simple2}), if a bin $(i,j)$ is not examined in Phase~$t+1$, then its least preferred applicant cannot be packed in phase~$t+1$ either. \qed

\end{proof}

We define an assignment $\mu^{\alpha}$ based on a number $\alpha \in [0,1)$ as follows. Assume that there is a line of height $\alpha$ ``cutting through'' all the bins horizontally. If an item $x_{ia}$ whose position in $i$'s bins intersects $\alpha$, applicant $a$ is assigned to institute $i$. In the case this cutting line of height $\alpha$ intersects two items in the same bin, we choose the item occupying the higher position. More precisely: 

\begin{quote} Given $\alpha \in [0,1)$, for each institute $i \in \I$, we define an assignment as follows: $\mu^{\alpha}(i) = \{a: 1-x(T(i,a)) \leq \alpha < 1- x(T(i,a)) + x_{ia}\}$. 
\end{quote}

\begin{theorem} The polytope determined by Constraints~(\ref{equ:lpFirst})-(\ref{equ:lpFifth})  is integral. 

\end{theorem}

\begin{proof} We generate uniformly at random a number $\alpha \in [0,1)$ and use it to define an assignment $\mu^{\alpha}$. To facilitate the discussion, we choose the largest $\alpha' \leq \alpha$ so that $\mu^{\alpha'} = \mu^{\alpha}$. Intuitively, this can be regarded as lowering the cutting line from $\alpha$ to $\alpha'$ without modifying the assignment, and $1-\alpha'$ is exactly the maximum available space in the beginning of a certain phase $l$ during the execution  of our packing algorithm. Note that the assignment $\mu^{\alpha}$ is then equivalent to giving those applicants (items) on top of institute $i$'s bins to $i$ at the end of phase $l$. 

We now argue that $\mu^{\alpha}$ is a stable matching. First, it is a matching by Theorem~\ref{thm:packing}\textrm{(vi)}. The matching respects the quota of all classes since Invariant~$C$ is maintained. What remains to be argued is the stability of $\mu^{\alpha}$. Suppose, for a contradiction, $(i,a^{\phi})$ is a blocking pair. We consider the possible cases. 

\begin{itemize}

\item Suppose that $x_{ia^{\phi}}>0$ and $x_{ia^{\phi}}$ is not put into the bins yet at the end of Phase $l$. Then by Theorem~\ref{thm:packing}\textrm{(iv)} and the definition of blocking pairs, $(i,a^{\phi})$ cannot block $\mu^{\alpha}$. 

\item Suppose that $x_{ia^{\phi}}>0$ and $x_{ia^{\phi}}$ is already put into the bins at the end of Phase $l$. If $\mu^{\alpha}(a^{\phi})=i$, there  is nothing to prove. So assume $\mu^{\alpha}(a^{\phi})\neq i$ and this means that the item $x_{ia^{\phi}}$ is ``buried'' under some other item on top of some of $i$'s bins at the end of Phase $l$. Then by Theorem~\ref{thm:packing}\textrm{(v)}, $a^{\phi}$ is assigned to some other institute ranking higher than $i$, contradicting the assumption that $(i,a^{\phi})$ is a blocking pair. 

\item Suppose that $x_{ia^{\phi}}=0$. There are two subcases.

\begin{itemize}

\item Suppose that for each of the classes $C^i_{a^{\phi}t} \in a^{\phi}(\C(i))$, $|\mu^{\alpha}(i) \cap C^i_{a^{\phi}t}| < q^{+}(C^i_{a^{\phi}t})$. Then we can form a new feasible tuple $\mu^{\alpha}(i)|a^{\phi}$. It can be inferred from the definition of the shaft that $x(S(\mu^{\alpha}(i)|a^{\phi})) \leq x(S(\mu^{\alpha}(i))$. Moreover, by Theorem~\ref{thm:packing}\textrm{(iii)}, we have $x(K(i,S(\mu^{\alpha}(i))) = |E_i(x)|$. Now by Constraint~(\ref{equ:lpThird}),

\begin{eqnarray*} |E_i(x)|+1 & \leq &  x(K(i,S(\mu^{\alpha}(i)|a^{\phi}))) \\
         & \leq & x(S(\mu^{\alpha}(i)) + x(T(i,a^{\phi})\backslash \{(i,a^{\phi})\})+ 
         \sum_{a \in \mu^{\alpha}} x(T(i,a)\backslash \{(i,a)\}) \\
        & = & x(K(i,S(\mu^{\alpha}(i)))) +  x(T(i,a^{\phi})\backslash \{(i,a^{\phi})\}) \\
        & = & |E_i(x)| + x(T(i,a^{\phi})\backslash \{(i,a^{\phi})\}).
\end{eqnarray*} 

As a result, $x(T(i,a^{\phi})\backslash \{(i,a^{\phi})\}) =1$, implying that $\mu^{\alpha}(a^{\phi}) 
\succ_{a^{\phi}} i$, a contradiction to the assumption that $(i,a)$ blocks $\mu^{\alpha}$.   

\item Suppose that there exists a class $C^i_{a^{\phi}l^{\phi}} \in a^{\phi}(\C(i))$ for which 
$|\mu^{\alpha}(i) \cap C^i_{a^{\phi}l^{\phi}}| = q^{+}(C^i_{a^{\phi}l^{\phi}})$. Let $C^i_{a^{\phi}l^{\phi}}$ be the smallest such class. By definition of blocking pairs, there must exist an applicant $a^{\dagger} \in \mu^{\alpha}(i) \cap C^i_{a^{\phi}l^{\phi}}$ who ranks lower than $a^{\phi}$. Choose $a^{\dagger}$ to be the lowest ranking such applicant in $\mu^{\alpha}(i)$. We make the following critical observation: 

\begin{equation} 
x(S(\mu^{\alpha}(i)|^{a^{\dagger}}a^{\phi})) \leq x(S(\mu^{\alpha}(i))) - x_{ia^{\dagger}}.
\label{equ:criticalObservation}
\end{equation}

To see this, we first argue that given an item $x_{ia} >0$, if it does not contribute to the shaft $S(\mu^{\phi}(i))$, then it cannot contribute to shaft 
$S(\mu^{\alpha}(i)|^{a^{\dagger}}a^{\phi})$ either. It is trivial if $a \succ_i a^{\dagger}$. So assume that $a^{\dagger} \succ_i a$.
First suppose that $a \not \in C^i_{a^{\phi}l^{\phi}}$. 
Then given any class $C^i_{at} \in a(\C(i))$, 
$|\mu^{\alpha}(i) \cap C^i_{at} \cap \LL^i_{\succ a}| = 
|\mu^{\alpha}(i)|^{a^{\dagger}}a^{\phi} \cap C^i_{at} \cap \LL^i_{\succ a}|$, 
and Theorem~\ref{thm:packing}\textrm{(iv)} states that 
there is a class $C^i_{al} \in a(\C(i))$ such that $|\mu^{\alpha}(i) \cap C^i_{al} \cap \LL^i_{\succ a}| = 
q^{+}(C^i_{al})$. Secondly suppose that $a \in C^i_{a^{\phi}l^{\phi}}$. 
Observe that $q^{+}(C^i_{a^{\phi}l^{\phi}}) = |\mu^{\alpha}(i)|^{a^{\dagger}}a^{\phi} \cap C^i_{a^{\phi}l^{\phi}}\cap \LL^i_{\succ a^{\dagger}}|= 
 |\mu^{\phi}(i)|^{a^{\dagger}}a^{\phi} \cap C^i_{a^{\phi}l^{\phi}}\cap 
\LL^i_{\succ a}|$ (the first equality follows from the choice of $a^{\dagger}$). 
In both cases, we conclude that $x_{ia}$ cannot contribute to the 
shaft $S(\mu^{\phi}(i)|^{a^{\dagger}}a^{\phi})$. The term $x_{ia^{\dagger}}$ does not contribute to the shaft 
$S(\mu^{\phi}(i)|^{a^{\dagger}}a^{\phi})$ by the same argument. 
Now using Constraint~(\ref{equ:lpThird}), Theorem~\ref{thm:packing}\textrm{(iii)}, and Inequality~(\ref{equ:criticalObservation}), we have 

\begin{eqnarray*} 
|E_i(x)| &\leq &  x(K(i,S(\mu^{\alpha}(i)|^{a^{\dagger}}a^{\phi}))) \\
& \leq & x(S(\mu^{\alpha}(i))) - x_{ia^{\dagger}} + x(T(i,a^{\phi}) \backslash \{(i,a^{\phi})\})+ 
\sum_{a \in \mu^{\alpha}(i) \backslash \{a^{\dagger}\}} x(T(i,a) \backslash \{(i,a)\})) \\
& = & |E_i(x)| - x(T(i,a^{\dagger})) + x(T(i,a^{\phi})). \mbox{ \hspace*{0.25in} (Note that $x_{ia^{\phi}}=0$).}
\end{eqnarray*}

Therefore, 

$$x(T(i,a^{\phi})) \geq x(T(i,a^{\dagger})) \geq 1 - \alpha' \geq 1 - \alpha.$$ 

So $\mu^{\alpha}(a^{\phi}) \succ_{a^{\phi}} i$, again a contradiction to the assumption that 
$(i,a^{\phi})$ blocks $\mu^{\alpha}$. 

So we have established that the generated assignment $\mu^{\alpha}$ is a stable matching. Now the remaining proof is the same as in~\cite{Teo:2001a}. Assume that $\mu^{\alpha}(i,a)=1$ if and only if applicant $a$ is assigned to institute $i$ under $\mu^{\alpha}$. Then 

$$Exp[\mu^{\alpha}(i,a)] = x_{ia}.$$ 

Then $x_{ia} = \int_{0}^{1}\mu^{\alpha}(i,a)d \alpha$ and $x$ can be written as a convex combination of $\mu^{\alpha}$ as $\alpha$ varies over the interval $[0,1)$. The integrality of the polytope thus follows. \qed
\end{itemize}

\end{itemize}
\end{proof}

\subsection{Optimal Stable Matching}
\label{sec:optimal}
Since our polytope is integral, we can write suitable objective functions to target for various optimal stable matchings using Ellipsoid algorithm~\cite{Grotschel:1981}. As  the proposed LP has an exponential number of constraints, we also design a separation oracle to get a polynomial time algorithm. The basic idea of our oracle is based on dynamic programming. 

\subsection{Median-Choice Stable Matching} 
\label{sec:median}
An application of our polyhedral result is the following.

\begin{theorem} Suppose that in the given instance, all classifications are laminar families and there is no lower bound, $q^{-}(C^i_j)=0$ for any class $C^i_j$. Let $\mu_1$, $\mu_2$, $\cdots$, $\mu_k$ be stable matchings. If we assign every applicant to his median choice among all the $k$ matchings, the outcome is a stable matching. 
\label{thm:fair}
\end{theorem}

\begin{proof} Let $x_{\mu_t}$ be the solution based
 on $\mu_t$ for any $1 \leq t \leq k$ and apply our packing algorithm on the fractional solution $x= \frac{ \sum_{t=1}^{k} x_{\mu_t}}{k}$. Then let $\alpha=0.5$ and $\mu^{0.5}$ be the stable matching resulted from the cutting line of height $\alpha=0.5$. 
We make the following observation based on Theorem~\ref{thm:packing}:

\begin{quote} Suppose that applicant $a$ is matched under $x$ and those institutes with which he is matched are $i_1$, $i_2$, $\cdots$, $i_{k'}$, ordered based on their rankings on $a$'s preference list. Assume that he is matched to $i_{t}$ $n_{t}$ times among the $k$ given stable matchings. 
At the termination of the packing algorithm, each of the items $x_{i_{l}a}$, $1 \leq l \leq k'$, appears in institute $i_l$'s bins and its position is from 
$\sum_{t=1}^{l-1}\frac{n_{t}}{k}$ to $\sum_{t=1}^{l}\frac{n_{t}}{k}$. 
\end{quote}

Now $\mu^{0.5}$ gives every applicant his median choice follows easily from the above observation. \qed

\end{proof}

Using similar ideas, we can show that an applicant-optimal stable matching must be institute-(lexicographical)-pessimal and similarly an applicant-pessimal stable matching must be institute-(lexicographical)-optimal: by taking $x$ as the average of all stable matchings and consider the two matching $\mu^{\epsilon}$ and $\mu^{1-\epsilon}$ with arbitrary small $\epsilon>0$. 
Hence, it is tempting to conjecture that the median choice stable matching is also a lexicographical median outcome for the institutes. Somehow surprisingly, it turns out not to be the case and a counter-example can be found in the appendix. 

\subsection{Polytope for Many-to-Many ``Unclassified'' Stable Matching}

In the many-to-many stable matching problem, each entity $e \in \I \cup \A$ has a quota $Q(e) \in 
\mathbb{Z}^{+}$ and a preference over a subset of the other side. A matching $\mu$ is feasible if given any entity $e \in \I \cup \A$, (1) $|\mu(e)| \leq Q(e)$, and (2) $\mu(e)$ is a subset of the entities on $e'$s preference list. A feasible matching $\mu$ is stable if there is no blocking pair $(i,a)$, which means that $i$ prefers $a$ to one of the assignments $\mu(i)$, or if 
$|\mu(i)| < Q(i)$ and $a \not \in \mu(i)$; and similarly $a$ prefers $i$ to one of his assignments $\mu(a)$, or if $|\mu(a)| < Q(a)$ and $i \not \in \mu(a)$. 

We now transform the problem into (many-to-one) \textbf{LCSM}. For each applicant $a \in \A$, we create $Q(a)$ copies, each of which retains the original preference of $a$. All institutes replace the applicants by their clones on their lists. To break ties, all institutes rank the clones of the same applicant in an arbitrary but fixed manner. Finally, each institute treats the clones of the same applicant as a class with upper bound 1. It can be shown that the stable matchings in the original instance and in the transformed \textbf{LCSM} instance have a one-one correspondence. Thus, we can use Constraints~(\ref{equ:lpFirst})-(\ref{equ:lpFifth}) to describe the former\footnote{This cloning technique works for the many-to-one {\sc hospitals/residents} problem, i,.e, we can clone the hospitals and use the original one-one stable marriage polytope~\cite{Vande:1989} to describe that of the 
former~\cite{Sethuraman:2009,Sethuraman:2006a}. However, this trick does not work here  for many-to-many matching. The reason is that the same applicant cannot be assigned to an institute multiple times. And that is why our classifications help to resolve this problem.}. 

\section{Conclusion and Future Work}

In this paper, we introduce {\sc classified stable matching} and present a dichotomy theorem 
to draw a line between its polynomial solvability and NP-completeness. We also study the problem using the polyhedral approach and propose polynomial time algorithms to obtain various optimal matchings. 

We choose the terms ``institutes'' and ``applicants'' in our problem definition, 
instead of the more conventional hospitals and residents, for a reason. We are aware 
that in real-world academics, many departments not only have ranking over their 
job candidates but also classify them based on their research areas. 
When they make their hiring decision, they have to take the quota of the classes into 
consideration. And in fact, we were originally motivated by this common practice. 

{\sc classified stable matching} has happened in real world. 
In a hospitals/residents matching program in Scotland, certain hospitals declared that they
did not want more than one female physician. Roth~\cite{Roth:1991a} proposed an algorithm 
to show that stable matchings always exist. 

There are quite a few questions that remain open. The obvious one would be to write an LP to describe \textbf{LCSM} with both upper bounds and lower bounds. Even though we can obtain various optimal stable matchings,  the Ellipsoid algorithm can be inefficient. It would be nicer to have fast combinatorial algorithms. The \emph{rotation} structure of Gusfield and Irving~\cite{Gusfield:1989} seems the way to go.

\section*{Acknowledgments}

I thank Peter Winkler for suggesting to me the idea
of {\sc classified stable matching} and Alvin Roth and Jay Sethuraman for several fruitful discussions. The idea of reducing many-to-many stable matching to \textbf{LCSM} was suggested by Jay Sethuraman. The comments of SODA~10 reviewers have been extremely helpful. 

\bibliographystyle{plain}
\bibliography{ref_smp}

\newpage
\appendix

\section{An Example for Section~\ref{sec:ruralHospital}}

In contrast to the generalized rural hospitals theorem in \textbf{LCSM}, if some institutes use intersecting classes, stable matching sizes may differ. Figure~\ref{fig:differentSize} is an example. 

\begin{figure*}[h]\footnotesize
\hrule
\begin{tabbing}
\hspace{0pt}\=\hspace{100pt}\=\hspace{190pt}\=
\hspace{190pt}\=\hspace{30pt}\=\hspace{15pt}\\

\> Institute Preferences \>  Classifications     \> Quota   \\
\> $i_1\mbox{:} a_1 a_2 a_3     $  \> $C^1_1=\{a_1,a_2\}$, $C^1_2 = \{a_1,a_3\}$ \> 
$Q(i_1)=2$, $q^{+}(C^1_1)=1$, $q^{+}(C^1_2)=1$   \\
\> $i_2\mbox{:} a_2 a_1 a_3 a_4 $  \> $C^2_1 = \{a_2,a_1\}$, $C^2_2 = \{a_2,a_3\}$, $C^2_3 = \{a_2,a_4\}$ \> 
$Q(i_2)=2$, $q^{+}(C^2_1)=1$, $q^{+}(C^2_2)=1$, $q^{+}(C^2_3)=1$    \\\\\\

\> Applicant Preferences \\
\> $a_1\mbox{:} i_1 i_2  $ \\
\> $a_2\mbox{:} i_1 i_2  $   \\
\> $a_3\mbox{:} i_1 i_2$   \\
\> $a_4\mbox{:} i_2$ \\\\\\

\> Stable Matchings \\
\> $\mu_{A} = \{(i_1; a_1), (i_2;a_2)\}$ \\
\> $\mu_{B} = \{(i_1; a_2,a_3), (i_2;a_1,a_4)\}$ \\

\end{tabbing}
\vspace*{-0.35in}
\caption{An example of stable matchings of different sizes.}
\vspace*{0.03in}
\hrule
\label{fig:differentSize}
\end{figure*}

\section{Missing Proofs of Section~\ref{sec:np}} 

In this section, we prove Theorem~\ref{thm:notDownwardForstIsNPC}. 
We assume that the set of posets $\mathbb{P}=\{P_1,P_2,\cdots, P_k\}$ contains a poset which is not a downward forest. Moreover, we assume that there is no lower bound on the classes. 

Without loss of generality, we assume that $P_1$ is not a downward forest. Such a poset must 
have a ``V.'' By definition, there exists institute $i$ whose class inclusion 
poset $P(i)$ is isomorphic to $P_1$. This implies that institute $i$ must 
have two intersecting 
classes in $\C(i)$. In the following, we will present a reduction in which all institutes use \emph{at most} two classes (that can be intersecting). It is straightforward to use some dummy institutes and applicants to ``pad'' our reduction 
so that every poset $P_j \in \mathbb{P}$ is isomorphic to 
some class inclusion poset of the institutes in the derived instance. 
Our reduction is from {\sc one-in-three-sat}. We will use an instance 
in which there is no negative literal. (NP-completeness still holds under this 
restriction~\cite{Garey:1979}.) 

The overall goal is to design a reduction so that the 
derived $\mathbb{P}$-{\sc classified stable matching} instance 
allows a stable matching if and only if the given instance 
$\phi=c_1\wedge  c_2\wedge \cdots \wedge c_k$ is satisfiable. 
We will build a set of \emph{clause gadgets} to represent each clause $c_j$. 
For every pair of literals which belong to the same clause, we create a 
\emph{literal-pair} gadget. Such a gadget will guarantee that at most one literal 
it represents can be ``activated'' (set to TRUE). The clause gadget interacts with 
the literal-pair gadgets in such a way that if the clause is to be satisfied,  
exactly one literal it contains can be activated. 

\subsubsection*{Literal-Pair Gadget}

Suppose that $x^j_i$ and $x^j_{i'}$ both belong to the same clause $c_j$. We create
a gadget $\Upsilon^j_{i,i'}$ composed of
four applicants $\{a^j_{i,t}\}_{t=1}^{2} \cup \{a^j_{i',t}\}_{t=1}^{2}$ and two institutes $\{I^j_{i}, I^j_{i'}\}$ whose preferences and classifications are summarized below.

\begin{table}[h]\footnotesize
\begin{center}
 \begin{tabular}{ll|lll}
   $a^j_{i,1}$:& $I^j_{i} \succ \Gamma(a^j_{i,1}) \succ I^j_{i'}$  & $I^j_i$: & $a^j_{i,2}\succ a^j_{i,1} \succ
 a^j_{i',2} \succ a^j_{i',1} \succ \Psi(I^j_{i})$ & $C^{I^j_{i}}_1=\{a^j_{i,1},a^j_{i,2}\}$,
 $C^{I^j_{i}}_2=\{a^j_{i,1},a^j_{i',1}\}$   \\
  $a^j_{i,2}$:& $I^j_{i'} \succ I^j_{i} $ & & &$Q(I^j_i)=2$, $q^{+}(C^{I^j_{i}}_1)=1$, $q^{+}(C^{I^j_{i}}_2)=1$\\
$a^j_{i',1}$:& $I^j_{i} \succ \Gamma(a^j_{i',1}) \succ I^j_{i'}$  & $I^j_{i'}$: & $a^j_{i,1}\succ a^j_{i,2} \succ
 a^j_{i',1} \succ a^j_{i',2} $ & $C^{I^j_{i'}}_1=\{a^j_{i,1},a^j_{i,2}\}$
  \\
  $a^j_{i',2}$:& $I^j_{i'} \succ I^j_{i}$ & & & $Q(I^j_{i'})=2$, $q^{+}(C^{I^j_{i'}}_1)=1$\\
 \end{tabular}
\end{center}
\end{table}

We postpone the explanation of the $\Gamma$ and $\Psi$ functions for the time being. We first make the following claim.

\begin{quote} \textbf{Claim~$B$}: 
Suppose that in a stable matching $\mu$, the only possible assignments for
$\{a^j_{i,1},a^j_{i,2}, a^{j}_{i',1}, a^{j}_{i',2}\}$ are $\{I^{j}_{i},I^{j}_{i'}\}$.
 Then there can only be
three possible outcomes in $\mu$.

\begin{enumerate}

\item $\mu(a^j_{i,1})=I^j_i$,  $\mu(a^j_{i,2})=I^j_{i'}$,  $\mu(a^j_{i',1})=I^j_{i'}$,  $\mu(a^j_{i',2})=I^j_i$. (In
this case, we say $x_i$ is activated while $x_{i'}$ remains deactivated.)

\item $\mu(a^j_{i,1})=I^j_{i'}$,  $\mu(a^j_{i,2})=I^j_{i}$,  $\mu(a^j_{i',1})=I^j_{i}$,  $\mu(a^j_{i',2})=I^j_{i'}$. (In this case, we say $x_{i'}$ is activated while $x_{i}$ remains deactivated.)

\item $\mu(a^j_{i,1})=I^j_{i'}$,  $\mu(a^j_{i,2})=I^j_{i}$,  $\mu(a^j_{i',1})=I^j_{i'}$,  $\mu(a^j_{i',2})=I^j_i$. (In
this case, we say both $x_i$ and $x_{i'}$ remain deactivated.)

\end{enumerate}

\end{quote} 

Claim~$B$ can be easily verified.
Note that the case $\mu(a^j_{i,1})=I^j_i$,
$\mu(a^j_{i,2})=I^j_{i'}$,  $\mu(a^j_{i',1})=I^j_{i}$,  $\mu(a^j_{i',2})=I^j_{i'}$ will not happen due to the
quota $q^{+}(C^{I^j_{i}}_2)$. This case corresponds to the situation that
$x_i$ and $x_{i'}$ are both activated and is what we want to avoid.

We now explain how to realize the supposition in Claim~$B$ about the fixed potential
assignments for $\{a^j_{i,t}\}_{t=1}^{2} \cup \{a^j_{i',t}\}_{t=1}^{2}$ in a stable matching. It can be easily checked that if
$a^j_{i,1}$ is matched to some institute in $\Gamma(a^j_{i,1})$, or either of $\{a^j_{i,1}, a^j_{i,2}\}$ is unmatched; or if either of $\{a^j_{i',1}, a^j_{i',2}\}$ is unmatched, then 
there must exist a blocking group involving a subset of $\{I^j_{i},I^j_{i'}, 
\{a^j_{i,t}\}_{t=1}^2, \{a^j_{i',t}\}_{t=1}^{2}\}$.
However, the following matching $\mu^{\phi}$
can happen in which $a^j_{i',1}$ is matched to some institute in $\Gamma(a^j_{i',1})$ but there is no blocking group
: $\mu^{\phi}(a^j_{i,1})=I^j_i$, $\mu^{\phi}(a^j_{i,2})= \mu^{\phi}(a^j_{i',2})=I^j_{i'}$,
$\mu^{\phi}(a^j_{i',1}) \in \Gamma(a^j_{i',1})$.\footnote{It can be verified that if $a^j_{i,1}$ is matched to some institute in $\Gamma(a^j_{i',1})$, the above assignment is the only possibility that no blocking group arises.}

To prevent the above scenario from happening (i.e., we want $\mu^{\phi}$ to be unstable),
we introduce another gadget $\overline{\Upsilon}^j_{i}$, associated with
$I^j_{i}$, to guarantee a blocking group
will appear. We now list the preferences and classifications of the members of
$\overline{\Upsilon}^j_{i}$ below.

\begin{table}[h]\footnotesize
\begin{center}
 \begin{tabular}{ll|lll}
   $\overline{a}^j_{i,1}$:& $\overline{I}^j_{i,4} \succ
\overline{I}^j_{i,1} \succ
\overline{I}^j_{i,3} \succ
\overline{I}^j_{i,2} $  & $\overline{I}^j_{i,1}$: & $\overline{a}^j_{i,5} \succ
\overline{a}^j_{i,2} \succ
\overline{a}^j_{i,4} \succ
\overline{a}^j_{i,6} \succ
\overline{a}^j_{i,3} \succ
\overline{a}^j_{i,1} $
 & $Q(\overline{I}^j_{i,1})=2 $  \\

 $\overline{a}^j_{i,2}$:& $\overline{I}^j_{i,3} \succ
\overline{I}^j_{i,4} \succ
\overline{I}^j_{i,2} \succ
\overline{I}^j_{i,1} $  & $\overline{I}^j_{i,2}$: & $\overline{a}^j_{i,4} \succ
\overline{a}^j_{i,6} \succ
\overline{a}^j_{i,2} \succ
\overline{a}^j_{i,3} \succ
\overline{a}^j_{i,1} \succ
\overline{a}^j_{i,5}$  &  $C^{\overline{I}^j_{i,2}}_1=\{\overline{a}^j_{i,1},\overline{a}^j_{i,2}
, \overline{a}^j_{i,3}\}$, $C^{\overline{I}^j_{i,2}}_2=\{\overline{a}^j_{i,3},\overline{a}^j_{i,4}
, \overline{a}^j_{i,5}\}$  \\

$\overline{a}^j_{i,3}$:& $\overline{I}^j_{i,4} \succ
\overline{I}^j_{i,3} \succ
\overline{I}^j_{i,1} \succ
\overline{I}^j_{i,2} $  &
 & &  $Q(\overline{I}^j_{i,2})=2$, $q^{+}(C^{\overline{I}^j_{i,2}}_1)=1$, $q^{+}(C^{\overline{I}^j_{i,2}}_2)=1$  \\

 $\overline{a}^j_{i,4}$:& $\overline{I}^j_{i,4} \succ
\overline{I}^j_{i,1} \succ
\overline{I}^j_{i,2} \succ
\overline{I}^j_{i,3} $  & $\overline{I}^j_{i,3}$: & $\overline{a}^j_{i,4} \succ
\overline{a}^j_{i,5} \succ
\overline{a}^j_{i,6} \succ
\overline{a}^j_{i,3} \succ
\overline{a}^j_{i,1} \succ
\overline{a}^j_{i,2}$  &  $C^{\overline{I}^j_{i,3}}_1=\{\overline{a}^j_{i,1},\overline{a}^j_{i,2}
, \overline{a}^j_{i,3}\}$, $C^{\overline{I}^j_{i,3}}_2=\{\overline{a}^j_{i,3},\overline{a}^j_{i,4}
, \overline{a}^j_{i,5}\}$  \\

$\overline{a}^j_{i,5}$:& $\overline{I}^j_{i,2} \succ
\overline{I}^j_{i,4} \succ
\overline{I}^j_{i,3} \succ
\overline{I}^j_{i,1} $  &
 & &  $Q(\overline{I}^j_{i,3})=2$, $q^{+}(C^{\overline{I}^j_{i,3}}_1)=1$, $q^{+}(C^{\overline{I}^j_{i,3}}_2)=1$  \\
 $\overline{a}^j_{i,6}$:& $\overline{I}^j_{i,2} \succ
\overline{I}^j_{i,4} \succ
\overline{I}^j_{i,3} \succ
\overline{I}^j_{i,1} $  & $\overline{I}^j_{i,4}$: & $\overline{a}^j_{i,4} \succ
\overline{a}^j_{i,1} \succ
\overline{a}^j_{i,6} \succ
\overline{a}^j_{i,2} \succ
\overline{a}^j_{i,3} \succ
\overline{a}^j_{i,4}$  &  $C^{\overline{I}^j_{i,4}}_1=\{\overline{a}^j_{i,1},\overline{a}^j_{i,2}
, \overline{a}^j_{i,3}\}$, $C^{\overline{I}^j_{i,4}}_2=\{\overline{a}^j_{i,3},\overline{a}^j_{i,4}
, \overline{a}^j_{i,5}\}$  \\
& & &  & $Q(\overline{I}^j_{i,4})=2$, $q^{+}(C^{\overline{I}^j_{i,4}}_1)=1$, $q^{+}(C^{\overline{I}^j_{i,4}}_2)=1$

\end{tabular}
\end{center}
\end{table}

The above instance $\overline{\Upsilon}^{j}_i$ has
the following features, every one of which is crucial in our construction.

\begin{description}

\item [Feature~A] $\overline{\Upsilon}^j_{i}$ does not allow a stable matching; more importantly, if in the given matching
$\mu$, $|\mu(\overline{I}^{j}_{i,1})|<2$, then there exists at least one blocking group which is \emph{not}
of the form $(\overline{I}^j_{i,1};\overline{a}^j_{i,x},\overline{a}^{j}_{i,y})$, where
$\{\overline{a}^j_{i,x},\overline{a}^{j}_{i,y}\} \subset \{\overline{a}^j_{i,t}\}_{t=1}^{6}$.

\item [Feature~B] When the institute $\overline{I}^{j}_{i,1}$ is ``removed'' from the instance $\overline{\Upsilon}^j_{i}$
(i.e.  $\overline{I}^{j}_{i,1}$ is struck from the preferences
of the applicants $\{\overline{a}^j_{i,t}\}_{t=1}^{6}$),
there exists at least one stable matching.
For instance, the following matching $\mu$ is stable, if $\overline{I}^{j}_{i,1}$
is removed: $\mu(\overline{I}^j_{i,2}) = (\overline{a}^j_{i,6}, \overline{a}^j_{i,3})$,
$\mu(\overline{I}^j_{i,3}) = (\overline{a}^j_{i,5}, \overline{a}^j_{i,2})$,
$\mu(\overline{I}^j_{i,4}) = (\overline{a}^j_{i,4}, \overline{a}^j_{i,1})$.

\item [Feature~C] Each of the three institutes
$\{\overline{I}^j_{i,t}\}_{t=2}^{4}$ uses exactly two intersecting classes; more
importantly, institute $\overline{I}^{j}_{i,1}$ does not use any classification at all.

\end{description}

Our idea is to let the institute $I^j_{i}$ in the gadget $\Upsilon^{j}_{i,i'}$
``play the role'' of $\overline{I}^{j}_{i,1}$ in the gadget of $\overline{\Upsilon}^{j}_i$
and add the other members in $\overline{\Upsilon}^{j}_i$ into gadget $\Upsilon^j_{i,i'}$.
To be precise, let $\Psi(I^j_i) = \overline{a}^j_{i,5} \succ \overline{a}^j_{i,2} \succ
\overline{a}^j_{i,4} \succ
\overline{a}^j_{i,6} \succ
\overline{a}^j_{i,3} \succ
\overline{a}^j_{i,1} $ (the same preference list of $\overline{I}^j_{i,1}$) and add the
other members of $\overline{\Upsilon}^j_{i}$
into the gadget $\Upsilon^j_{i,i'}$
without modifying their preference lists (and classifications). We now explain
how the above features of $\overline{\Upsilon}^j_{i}$ help to realize the
supposition in Claim~$B$ about the potential assignments of the applicants $\{a^j_{i,t},a^j_{i',t}\}_{t=1}^2$ in a stable matching.

\begin{enumerate}

\item In a matching $\mu^{\phi}$, suppose that institute $I^j_{i}$ is only assigned $a^j_{i,1}$
while $a^j_{i',1}$ is assigned to some institutes in $\Gamma(a^j_{i',1})$ (the problematic case we discussed above). As a result,
institute $I^j_i$ can take one more applicant from the set $\{\overline{a}^j_{i,t}\}_{t=1}^{6}$. By Feature~$A$, there must exist a blocking group involving the members in
$\overline{\Upsilon}^j_i$. More importantly, this
blocking group need not be composed of $I^j_i$ and \emph{two} applicants from
$\{\overline{a}^j_{i,t}\}_{t=1}^{6}$.

\item In a matching $\mu^{\phi}$, suppose that institutes $I^j_i$ is assigned
two applicants from the set $\{a^j_{i,t} , a^j_{i',t}\}_{t=1}^2$. Then $\overline{I}^{j}_{i,1}$ can be regarded as being
removed from the instance $\overline{\Upsilon}^j_{i}$. And there exists a stable matching among the other members
of the instance $\overline{\Upsilon}^j_i$. This explains the necessity of Feature~$B$.

\item Finally, since $I^j_i$ already uses two intersecting classes, $\overline{I}^{j}_{i,1}$ should
not use any more classes. This explains the reason why Feature~$C$ is necessary.

\end{enumerate}
We have left the functions $\Gamma(a^j_{i,1})$ and $\Gamma(a^j_{i',1})$
unexplained so far. They contain institutes belonging to
the clauses gadgets, which will be the final component in our construction.

\subsubsection*{Clause Gadget}

Suppose that $c_j = x^j_{1} \vee x^j_{2} \vee x^j_{3}$. We create a clause gadget $\hat{\Upsilon}_{j}$
composed of two institutes $\{\hat{I}^j_t\}_{t=1}^2$
and six applicants $\{\hat{a}^j_{t}\}_{t=1}^{6}$. Their preferences and classifications are summarized below.

\begin{table}[h]\footnotesize
\begin{center}
 \begin{tabular}{ll|ll}
$\hat{a}^j_{1}$:& $\hat{I}^j_{2} \succ \hat{I}^j_1$ & $\hat{I}^j_1$: & $\hat{a}^j_{5} \succ \hat{a}^j_{1}
\succ \hat{a}^j_{2}
\succ \Lambda(x^j_{1})
\succ \hat{a}^j_{6}
\succ \Lambda(x^j_{2})
\succ \hat{a}^j_{3}
\succ \Lambda(x^j_{3})
\succ \hat{a}^j_{4}$ \\
$\hat{a}^j_{2}$:& $\hat{I}^j_{1} \succ \hat{2}^j_1$ & & $C^{\hat{I}^j_{1}}_1 = \{\hat{a}^j_{1},\hat{a}^j_{2},\hat{a}^j_{5},\hat{a}^j_{6},
\Lambda(x^j_{1}) \}$,  $C^{\hat{I}^j_{1}}_2  = \{\hat{a}^j_{2}, \hat{a}^j_{5}, \hat{a}^j_{6},\Lambda(x^j_{2})\}$  \\

$\hat{a}^j_{3}$:& $\hat{I}^j_{2} \succ \hat{I}^j_1$ & & $Q(\hat{I}^j_1) = 3$,
$q^{+}(C^{\hat{I}^j_{1}}_1) = 2 $,  $q^{+}(C^{\hat{I}^j_{1}}_2) = 2 $  \\

$\hat{a}^j_{4}$:& $\hat{I}^j_{1} \succ \hat{I}^j_2$ & $\hat{I}^j_2$:
 & $\hat{a}^j_{6} \succ \hat{a}^j_{2}
\succ \hat{a}^j_{1}
\succ \hat{a}^j_{5}
\succ \hat{a}^j_{4}
\succ \hat{a}^j_{3}$ \\
$\hat{a}^j_{5}$:& $\hat{I}^j_{2} \succ \hat{I}^j_1$ &
& $C^{\hat{I}^j_{2}}_1 = \{\hat{a}^j_{1},\hat{a}^j_{2},\hat{a}^j_{5},\hat{a}^j_{6}\}$,
$C^{\hat{I}^j_{2}}_2  = \{\hat{a}^j_{1}, \hat{a}^j_{4}, \hat{a}^j_{6}\}$.  \\

$\hat{a}^j_{6}$:& $\hat{I}^j_{1} \succ \hat{I}^j_2$ & & $Q(\hat{I}^j_2) = 3$,
$q^{+}(C^{\hat{I}^j_{2}}_1) = 2 $,  $q^{+}(C^{\hat{I}^j_{2}}_2) = 2 $
\end{tabular}
\end{center}
\end{table}

We now explain how the $\Lambda$ functions in the
clause gadgets interact with the $\Gamma$ functions in the
literal-pair gadgets. The former is composed of applicants in the literal-pair gadgets while
the latter is composed of institutes in the clause gadgets. Our intuition is that
the only possible stable matchings in the clause gadgets will enforce
exactly one of its three literals to be activated.
To be precise, let $\pi(X)$ denote an arbitrary
order among the elements in the set $X$. Then:

\begin{eqnarray*}
\Lambda(x^j_t)=\pi(\{a^{j'}_{i,1}|\mbox{ if } a^{j'}_{i,1} \in \Upsilon^{j'}_{i,i'}
\mbox{ and } x^{j'}_i = x^j_t\}), \mbox{ for $t \in \{1,2,3\}.$} \\
\Gamma(a^j_{i,1}) = \pi(\{\hat{I}^{j'}_1| \mbox{ if clause $c_{j'}=x^{j'}_1 \vee
x^{j'}_2 \vee x^{j'}_3$ contains the literal $x^j_i$.}\})
\end{eqnarray*}

\begin{quote}
\textbf{Claim~$C$} Suppose there exists a stable matching $\mu$ in the derived instance. Then
there can be only three possible outcomes for the members in the clause gadget
$\hat{\Upsilon}_j$.

\begin{enumerate}

\item $\mu(\hat{I}^j_1)= (\hat{a}^{j}_5, \hat{a}^{j}_1, \hat{a}^{j}_3)$,
$\mu(\hat{I}^j_2)=(\hat{a}^{j}_6, \hat{a}^{j}_2, \hat{a}^{j}_4)$. Moreover, in this
case, $x^j_2$ must be activated while $x^j_1$ and $x^j_3$ must remain deactivated.

\item $\mu(\hat{I}^j_1)= (\hat{a}^{j}_5, \hat{a}^{j}_2, \hat{a}^{j}_4)$,
$\mu(\hat{I}^j_2)=(\hat{a}^{j}_6, \hat{a}^{j}_1, \hat{a}^{j}_5)$. Moreover, in this
case, $x^j_3$ must be activated while $x^j_1$ and $x^j_2$ must remain deactivated.

\item $\mu(\hat{I}^j_1)= (\hat{a}^{j}_2, \hat{a}^{j}_6, \hat{a}^{j}_3)$,
$\mu(\hat{I}^j_2)=(\hat{a}^{j}_1, \hat{a}^{j}_5, \hat{a}^{j}_4)$. Moreover, in this
case, $x^j_1$ must be activated while $x^j_2$ and $x^j_3$ must remain deactivated.

\end{enumerate}

\end{quote}

\textbf{Proof of Claim~$C$.} We have argued previously that in a stable matching $\mu$, the applicants in the
literal-pair gadgets cannot be matched to the institutes in the clause gadget. So both
$\hat{I}^j_1$ and $\hat{I}^j_2$ can only have applicants from the set 
$\{\hat{a}^j_t\}_{t=1}^6$ in $\mu$.
We first make the following observation.

\begin{quote}
In a stable matching $\mu$, one of the applicants in the following three
pairs $\{\hat{a}^j_1, \hat{a}^j_2\}$,  $\{\hat{a}^j_3, \hat{a}^j_4\}$,
$\{\hat{a}^j_5, \hat{a}^j_6\}$ must be assigned to $\hat{I}^j_1$; while
the other applicant in the above three pairs must be assigned to $\hat{I}^j_2$.
\end{quote}

The above observation can be easily verified. So there are eight possible combinations in $\mu$.
Except the three combinations listed in the claim, the other five will cause
blocking groups to arise.

\begin{enumerate}

\item $\mu(\hat{I}^j_1)=  (\hat{a}^{j}_1, \hat{a}^{j}_6, \hat{a}^{j}_3)$,
$\mu(\hat{I}^j_2)=(\hat{a}^{j}_2, \hat{a}^{j}_5, \hat{a}^{j}_4)$. Then
$(\hat{I}^j_1;\mu(\hat{I}^j_1)|^{\hat{a}^j_2} \hat{a}^{j}_6)$ blocks $\mu$.

\item $\mu(\hat{I}^j_1)=  (\hat{a}^{j}_1, \hat{a}^{j}_6, \hat{a}^{j}_4)$,
$\mu(\hat{I}^j_2)=(\hat{a}^{j}_2, \hat{a}^{j}_5, \hat{a}^{j}_3)$. Then again
$(\hat{I}^j_1; \mu(\hat{I}^j_1)|^{\hat{a}^j_2} \hat{a}^{j}_6)$ blocks $\mu$.

\item $\mu(\hat{I}^j_1)=  (\hat{a}^{j}_5, \hat{a}^{j}_2, \hat{a}^{j}_3)$. Note
that the tuple $\mu(\hat{I}^j_2)=(\hat{a}^{j}_6, \hat{a}^{j}_1, \hat{a}^{j}_4)$ is not
feasible. So $\mu(\hat{I}^j_2)=(\hat{a}^{j}_6, \hat{a}^{j}_1)$. But then
$(\hat{I}^j_2, \mu(\hat{I}^j_2)|\hat{a}^j_3)$ blocks $\mu$.

\item $\mu(\hat{I}^j_1)=  (\hat{a}^{j}_5, \hat{a}^{j}_1, \hat{a}^{j}_4)$,
$\mu(\hat{I}^j_2)=(\hat{a}^{j}_6, \hat{a}^{j}_2, \hat{a}^{j}_3)$. In this case, institute
$\hat{I}^j_1$ can replace $\hat{a}^j_4$ with an applicant in $\Lambda(x^j_2)$
or an applicant in $\Lambda(x^j_3)$ to have a blocking group, unless
both $x^j_2$ and $x^j_3$ are activated. However, this is impossible, since we have a literal-pair
gadget $\Upsilon^{j}_{2,3}$ and by Claim~$B$, at most one
of $x^j_2$ and $x^j_3$ can be activated. So we have a contradiction.

\item  $\mu(\hat{I}^j_1)=  (\hat{a}^{j}_2, \hat{a}^{j}_6, \hat{a}^{j}_4)$,
$\mu(\hat{I}^j_2)=(\hat{a}^{j}_1, \hat{a}^{j}_5, \hat{a}^{j}_3)$. Then the existence
of a blocking group involving $\hat{I}^j_1$ and some applicants in $\Lambda(x^j_1)
\cup \Lambda(x^j_3)$ is guaranteed. The reason is similar to the previous case:
$x^j_1$ and $x^j_3$ cannot be both activated.
\end{enumerate}

Finally, we remark that the three possible outcomes in $\mu$ listed in the lemma will guarantee that exactly
one of the three literals in clause $c_j$ can be activated. The reason
is again the same as in the last two cases that we just explained. This completes the proof
of Claim~$C$. \qed

Now by Claim~$C$, we establish Theorem~\ref{thm:notDownwardForstIsNPC}

\section{Missing Proofs of Section~\ref{sec:polyhedral}} 
\setcounter{theorem}{16}
\begin{lemma} In \textbf{LCSM}, if there is no lower bound, i.e., given any class $C^i_j$, $q^{-}(C^i_j)=0$, then a stable matching as defined in Definition~\ref{def:first} can be equivalently defined as follows. A feasible matching $\mu$ is stable if and only if there is no blocking pair. A pair $(i,a)$ is blocking, given that $\mu(i) = (a_{i1}, a_{i2},\cdots, a_{ik})$, $k \leq Q(i)$, if 

\begin{itemize}
\item $i \succ_{a} \mu(a)$; 
\item for any class $C^i_{at} \in a(\C(i))$, $|\LL^i_{\succ a} \cap \mu(i) \cap C^i_{at}| < q^{+}(C^i_{at})$. 
\end{itemize}
\label{pro:blockingPair}
\end{lemma}

\begin{proof} If we have a blocking group $(i;g)$, institute $i$ and the highest ranking applicant in $g \backslash \mu(i)$ must be a blocking pair. Conversely, given a blocking pair $(i;a)$, assuming that $|\mu(i)| = Q(i)$ (the case that $|\mu(i)| < Q(i)$ follows a similar argument), we can form a blocking group $(i;\mu(i)|^{a^{\dagger}} a)$, where
$a^{\dagger}$ is chosen as follows: (1) if there exists a class $C^i_{at} \in a(\C(i))$ such that $|\mu(i) \cap C^i_{at}|= q^{+}(C^i_{at})$, choose the smallest such class $C^i_{at} \in a(\C(i))$ and let $a^{\dagger}$ be the lowest ranking applicant in $\mu(i) \cap C^i_{at}$; (2) otherwise, $a^{\dagger}$ is simply the lowest ranking applicant in $\mu(i)$. \qed

\end{proof}

\setcounter{theorem}{18}
\begin{lemma} 
Every stable matching solution $x$ satisfies the \emph{comb inequality} for any comb $K(i,S(A^i))$:
\begin{equation*}
x(K(i,S(A^i)) \equiv x(S(A^i)) + \sum_{a_j \in A^i} x(T(i,a_j)\backslash \{i, a_j\}) \geq |A^i|. \label{equ:stableConstraint}
\end{equation*}
\end{lemma}

We use the following notation to facilitate the proof. Give a tuple $A^i$, we define $y_{ia}$ as follows: 

$$y_{ia} = \left\{ \begin{array}{ll}
   1 & \mbox{ either } a \in A^i, x(T(i,a)) = 1; \mbox { or } a \not \in A^i, x_{ia}=1, \mbox{ and } (i,a) \in S(A^i);\\
   0 & \mbox{ o.w. } 
   \end{array}
\right. 
$$

Let $y(C^i_j) = \sum_{a \in \LL^i \cap C^i_j} y_{ia}$. This quantity indicates how much a class $C^i_j$ contributes to the comb value $x(K(i,S(A^i)))$. Thus, if $U$ is a set of classes in $\C(i)$ partitioning $\LL^i$, then $x(K(i,S(A^i)))  = \sum_{C^i_j \in U} y(C^i_j)$. 

\begin{proof} We prove by showing that if $x(K(i,S(A^i))) < |A^i|$, there exists a blocking pair $(i,a^{\dagger})$, where $a^{\dagger} \in A^i$. We proceed by contradiction. 
First note that there exists a non-empty subset $G \subseteq A^i$ of applicants $a$ for whom $x(T(i,a))=0$, otherwise, $x(K(i,S(A^i))) \geq |A^i|$, an immediate contradiction. For each applicant $a \in G$, there must exist a class $C^i_{al} \in a(\C(i))$ for which $\sum_{a' \in \LL^i_{\succ a} \cap C^i_{al}} x_{ia'} = q^{+}(C^i_{al})$, otherwise, $(i,a)$ is a blocking pair and we are done. Now for each applicant $a \in G$, choose the smallest class $C^i_{al}$  
for which $\sum_{a' \in \LL^i_{\succ a} \cap C^i_{al}} x_{ia'} = q^{+}(C^i_{al})$ and denote this class as $\overline{C}_a$. We introduce a procedure to organize a set $U$ of disjoint classes. 

\begin{quote} Let $G$ be composed of $a_1$, $a_2$, $\cdots$, $a_{|G|}$ ordered based on their decreasing rankings on $\LL^i$\\
\textbf{For} $i=1$ \textbf{To} $|G|$ \\
\hspace*{0.2in} \textbf{if} $a_i \in C \in U$, \textbf{then} do nothing\\
\hspace*{0.2in} \textbf{else} $U := U \backslash \{C|C \in U, C \subset \overline{C}_{a_l}\}$ 
//$\overline{C}_{a_l}$ may be a superclass of some classes in $U$\\
\hspace*{0.53in}                       $U := U \cup \{\overline{C}_{a_l}\}$. // adding $\overline{C}_{a_l}$ into $U$
\end{quote}

\begin{quote} \textbf{Claim} The output $U$ from the above procedure comprises of a disjoint set of classes containing all applicants in $G$, and for each class $C^i_j \in U$, $y(C^i_{j}) \geq q^{+}(C^i_j).$ 

\end{quote}

We will prove the claim shortly. Now 

\begin{eqnarray*}
x(K(i, S(A^i))) &= & \sum_{C^i_j \in U}y(C^i_j) +  |A^i \backslash \{\cup_{C^i_j \in U}C^i_j\}| 
\geq \sum_{C^i_j \in U}q^{+}(C^i_j) + |A^i \backslash \{\cup_{C^i_j \in U}C^i_j\}| \geq |A^i|, 
\end{eqnarray*}

\noindent a contradiction. \qed

\textbf{Proof of the Claim.} It is easy to see that the classes in $U$ are disjoint and contain all applicants in $G$. Below we show that during the execution of the procedure, if $C^i_j \in U$, 
then $y(C^i_j) \geq q^{+}(C^i_j)$. We proceed by induction on the number of times $U$ is updated. In  the base case $U$ is an empty set so there is nothing to prove. 

For the induction step, assume that $a_l$ is being examined and
 $\overline{C}_{a_l}$ is about to be added into $U$. Observe that even though $\sum_{a \in \LL^i_{\succ a_l} 
\cap \overline{C}_{a_l}}x_{ia} = q^{+}(\overline{C}_{a_l})$, there is no guarantee that if $x_{ia}=1$, then $y_{ia}=1$ for each $a \in \LL^i_{\succ a_l} \cap \overline{C}_{a_l}$. The reason is that there may exist some class $C^i_j \in a(\C(i))$ for which $|A^i \cap C^i_j \cap \LL^i_{\succ a}|= q^{+}(C^i_j)$ and $a \not \in A^i$. Then $(i,a)$ is not part of the shaft $x(S(A^i))$ and $y_{ia}=0$. 

To deal with the above situation, we need to do some case analysis. Let $B$ be the set of subclasses $C^i_j$ of $\overline{C}_{a_l}$ for which 
$|A^i \cap C^i_j \cap \LL^i_{\succ a_l}| = q^{+}(C^i_j)$. 
Choose $D$ to be the subclasses of $\overline{C}_{a_l}$ so that 
$\Re(B \cup U) \cup D$ partitions $\overline{C}_{a_l}$. We make three observations below. 

\begin{quote} \textrm{(i)} for each class $C^i_j \in \Re( B \cup U)$ and $C^i_j \in U$, $y(C^i_j) \geq q^{+}(C^i_j) \geq \sum_{a \in \LL^i_{a_l} \cap C^i_j}x_{ia}$.\\
\textrm{(ii)} for each class $C^i_j \in D$, if $a \in \LL^i_{\succ a_l} \cap C^i_j$ and $x_{ia}=1$, then $y_{ia}=1$.\\
\textrm{(iii)} for each class $C^i_j \in \Re( B \cup U)$ and $C^i_j \not \in U$, then for each applicant $a \in \LL^i_{\succ a_l} \cap C^i_j \cap A^i$, either $a \in G$ and 
$a \in C\in U$, or that $a \not \in G$ (implying that $x(T(i,a))=1$). Moreover,  
$y(C^i_j) \geq \sum_{a \in \LL^i_{\succ a_l} \cap C^i_j}x_{ia}$
\end{quote}

\textrm{(i)} is because of the induction hypothesis and the feasibility assumption of $x$. 
\textrm{(ii)} follows from the fact that $a$ ranks higher than $a_l$ and the way we define a class in $D$. For \textrm{(iii)}, first notice that if $C^i_j \in \Re( B \cup U)$ and $C^i_j \not \in U$, then such a class $C^i_j$ must  be part of $\Re(B)$ and $C^i_j$ may contain some classes in $U$. Now suppose that $a_i \in G \cap \LL^i_{\succ a_l}$ but does not belong to any class in $U$. Then our procedure would have added the class $\overline{C}_{a_i}$ into $U$ before examining $a_l$, a contradiction. To see the last statement of \textrm{(iii)}, let $G'$ be set of applicants in 
$\LL^i_{\succ a_l} \cap C^i_j \cap A^i$ who do not belong to any classes in $U$. Then 

$$y(C^i_j) \geq \sum_{C^i_{k} \in U, C^i_k \subset C^i_j}y(C^i_k) + |G'| \geq 
\sum_{C^i_{k} \in U, C^i_k \subset C^i_j}q^{+}(C^i_{k}) + |G'| \geq q^{+}(C^i_j) \geq
\sum_{a \in \LL^i_{\succ a_l} \cap C^i_j}x_{ia},$$ 

\noindent where the first inequality follows from the first part of \textrm{(iii)}, the second inequality the induction hypothesis, the third the fact that $C^i_j \in \Re(B)$ (thus $|\LL^i_{\succ a_l} \cap C^i_j \cap A^i| = q^{+}(C^i_j)$), and the fourth the feasibility assumption of $x$. 

Now combining all the three observations, we conclude that  

$$y(\overline{C}_{a_l})=  \sum_{C^i_j \in \Re(B \cup U)} y(C^i_j) 
+  \sum_{C^i_k \in D} y(C^i_j) \geq \sum_{C^i_k \in \Re(B_l \cup U ) \cup D_l} \sum_{a \in \LL^i_{\succ a_l} \cap C^i_k}x_{ia}=q^{+}(C^i_j),$$  

\noindent and the induction step is completed. \qed

\end{proof}

\setcounter{theorem}{19}
\begin{lemma} Every stable matching solution $x$ satisfies the following inequality for any class-tuple $t^i_j$: 
\begin{equation*}
\sum_{a_{ij} \in t^i_j} x(T(i,a_{ij}) \backslash \{i,a_{ij}\}) \geq \sum_{a \in C^i_j \cap \LL^i_{\prec t^i_j}}x_{ia} \mbox{\hspace*{0.2in}  (*)}
\end{equation*}
\label{pro:newConstraint}
\end{lemma}

\begin{proof} We prove by contradiction. Suppose that in a given class-tuple $t^i_j$ (*) does not hold. We will show that we can find a blocking pair $(i,a^{\dagger})$, where $a^{\dagger} \in t^i_j$. Let the set of applicants $a \in t^i_j$ with $x(T(i,a))=0$ be  
$G$, $\alpha = \sum_{a' \in \LL^i_{\prec t^i_j} \cap C^i_j} x_{ia'} > 0$, and 
$\beta = \sum_{a' \in t^i_j}x_{ia'}$. By assumption, at most $\alpha-1$ applicants $a \in t^i_j$ have $x(T(i,a)\backslash \{(i,a)\}) = 1$. Thus, 
\begin{equation}
|G| \geq q^{+}(C^i_j) -\beta -\alpha+1. \label{equ:alphaBeta}
\end{equation}


\begin{quote} \textbf{Claim}: At least one applicant $a^{\dagger} \in G$ belongs to a sequence 
of classes $C^i_{a^{\dagger}t} \in a^{\dagger}(\C(i)) $ such that 
if $C^i_{a^{\dagger}t} \subseteq C^i_j$, then 
$\sum_{a' \in \LL^i_{\succ a^{\dagger}} \cap C^i_{a^{\dagger}t}} x_{ia'} < q^{+}(C^i_{a^{\dagger}t})$. 
\end{quote}

We will prove the claim shortly.  Observe that given any class $C^i_k \supset C^i_j$, $\sum_{a' \in \LL^i_{\succ a^{\dagger}} \cap C^i_k} x_{ia'} < q^{+}(C^i_k)$: as $\alpha > 0$, some applicant $a^{\phi}\in C^i_k$ ranking lower than $a^{\dagger}$ has $x_{ia^{\phi}}=1$ and 
Constraint~(\ref{equ:lpSecond}) enforces that $\sum_{a' \in \LL^i \cap C^i_k}x_{ia'} \leq q^{+}(C^i_k)$. Combining the above facts, we conclude that $(i, a^{\dagger})$ is a blocking pair. \qed

\end{proof}

\textbf{Proof of the Claim}. We prove by contradiction. Suppose that for every applicant $a \in G$, 
there exists some class $C^i_{at} \in a(\C(i))$, $C^i_{at} \subseteq C^i_j$, and 
$\sum_{a' \in \L^i_{\succ a} \cap C^i_{at}} x_{ia'} = q^{+}(C^i_{at})$. 

Let $B$ be the set of classes $C^i_k \subseteq C^i_j$ such that 
$C^i_k$ contains an applicant $a \in G$ and $\sum_{a' \in \LL^i_{\succ a} \cap C^i_k} x_{ia'} = q^{+}(C^i_k)$  (which then will equal $\sum_{a' \in \LL^i_{\succeq t^i_j} \cap C^i_k}x_{ia'}$ due to Constraint~(\ref{equ:lpSecond})). For each class $C^i_k \in \Re(B)$, 

\begin{equation} \sum_{a \in \LL^i_{\succeq t^i_j} \cap C^i_k}x_{ia} 
= q^{+}(C^i_k) \geq |t^i_j \cap C^i_k| = \sum_{a' \in \LL^i_{\succeq t^i_j} \cap t^i_j \cap C^i_k}x_{ia'} + |G \cap C^i_k|,
\label{equ:CountingInPropositionTwo}
\end{equation}

\noindent where the first inequality follows from the definition of the class-tuple. 
Now we have 

$$q^{+}(C^i_j) -\alpha - \beta \geq 
\sum_{C^i_k \in \Re(B)} \sum_{a' \in (\LL^i_{\succ a^{\ddag}} \cap C^i_k) \backslash t^i_j}x_{ia'} \geq \sum_{C^i_k \in \Re(B)}|G \cap C^i_k| = |G| \geq q^{+}(C^i_j)-\alpha-\beta+1, 
$$

\noindent a contradiction. 
Note that the first inequality follows from Constraint~(\ref{equ:lpSecond}), the second inequality  from~(\ref{equ:CountingInPropositionTwo}), the equality right after is because every applicant in $G$ belongs to some class in $B$, and the last inequality is due to~(\ref{equ:alphaBeta}). \qed

\section{Separation Oracle in Section~\ref{sec:optimal}} 

It is clear that Constraints~(\ref{equ:lpFirst})(\ref{equ:lpSecond})(\ref{equ:lpFourth}) can be separated in polynomial time. So we assume that $x$ satisfies these constraints and focus on finding a violated Constraint~(\ref{equ:lpThird}) and/or Constraint (\ref{equ:lpFifth}). 

\subsubsection*{Separating Constraint~(\ref{equ:lpThird})}

We first make an observation. For each institute $i$, it suffices to check whether all the combs with exactly $Q(i)$ teeth satisfy Constraint~(\ref{equ:lpThird}). To see this, suppose that there is a feasible tuple $\overline{A}^i$ with less than $Q(i)$ applicants and $x(K(i,S(\overline{A}^i))) < |\overline{A}^i|$. Then we can add suitable applicants into $\overline{A}^i$ to get a feasible tuple $A^i$ with exactly $Q(i)$ applicants. Noticing that $x(S(A^i)) \leq x(S(\overline{A}^i))$, we have 

\begin{eqnarray*}
x(K(i,S(A^i))) &\leq & x(S(\overline{A}^i)) + 
\sum_{a \in \overline{A}^i}x(T(i,a)\backslash \{(i,a)\}) + 
 \sum_{a \in A^i \backslash \overline{A}^i}x(T(i,a)\backslash \{(i,a)\}) \\
& <   & |\overline{A}^i| +  \sum_{a \in A^i \backslash \overline{A}^i}
x(T(i,a)\backslash \{(i,a)\})  \\
&\leq & |\overline{A}^i| + 
|A^i| - |\overline{A}^i| \\
&= & |A^i|,
\end{eqnarray*}

\noindent where the last inequality follows from our assumption that $x$ satisfies Constraint~(\ref{equ:lpFirst}). 

To illustrate our idea, we first explain how to deal with the case that the original classification $\C(i)$ is just a partition over $\LL^i$ (before we add the pseudo root class $C^i_{\sharp}$). We want to find out the tuple $A^i$ of length $Q(i)$, whose lowest ranking applicant is $a^{\dagger}$, which gives the smallest $x(K(i,S(A^i)))$. If we have this information for all possible $a^{\dagger}$, we are done. Note that because of our previous discussion, if there is no feasible tuple of length $Q(i)$ whose lowest ranking applicant is $a^{\dagger}$, we can ignore those cases. 

Our main idea is to decompose the value of $x(K(i, S(A^i)))$ based on the classes and use dynamic programming to find out the combinations of the tooth-applicants that give the smallest comb values. More precisely, 

\setcounter{theorem}{26}
\begin{definition} Assume that $A^i_j \subseteq C^i_j$, $0 \leq |A^i_j| \leq q^{+}(C^i_j)$, and all applicants in $A^i_j$ rank higher than $a^{\dagger}$. Let 
  
\begin{eqnarray*}
x(A^i_j, a^{\dagger}) = \sum_{a \in 
\LL^i_{\succ a^{\dagger}} \cap C^i_j, (i,a) \in S(A^i_j)}x_{ia}+ \sum_{a \in A^i_j} x(T(i,a)\backslash \{(i,a)\})\\
Z(C^i_j, s_j, a^{\dagger}) = min_{A^i_j:A^i_j \subseteq C^i_j, |A^i_j|=s_j} 
x(A^i_j, a^{\dagger}).
\end{eqnarray*}
\label{def:so}
\end{definition}

Note that this definition requires that if $x_{ia}$ contributes to $x(A^i_j, a^{\dagger})$, then $a$ has to rank higher than $a^{\dagger}$, belongs to $C^i_j$, and the $(i,a)$ is part of the shaft $S(A^i_j)$. 

Suppose that we have properly stored all the possible values of $Z(C^i_j, s_j, a^{\dagger})$ and assume that $a^{\dagger} \in C^i_{j'}$. Then for each class $C^i_j \neq C^i_{j'}$, assume that $0 \leq s_j \leq q^{+}(C^i_j)$ and for class $C^i_{j'}$, $0 \leq s_{j'} \leq q^{+}(C^i_{j'})-1$, then
 the tuple $A^i$ whose lowest ranking applicant is $a^{\dagger}$, that gives the smallest comb value is the following one:

$$x(K(i, S(A^i))) = x(T(i,a^{\dagger})) + min_{s_j: \sum_{C^i_j \in \C(i)} s_j = Q(i)-1} \sum_{C^i_j \in \C(i)} Z(C^i_j, s_j, a^{\dagger}).$$ 

The above quantity can be calculated using standard dynamic programming technique. So the question boils down to how to calculate $Z(C^i_j, s_j, a^{\dagger})$. There are two cases. 

\begin{itemize}

\item Suppose that $s_j < q^{+}(C^i_j)$. Observe that all the positive $x_{ia}$, where $a \in C^i_j \cap \LL^i_{\succ a^{\dagger}}$ will contribute to the value $x(A^i_j, a^{\dagger})$. 
So, to calculate $Z(C^i_j, s_j, a^{\dagger})$, we only need to find out the $s_j$ applicants $a$ in $C^i_j \cap \LL^i_{\succ a^{\dagger}}$ with the smallest values $x(T(i,a)\backslash \{(i,a)\})$. This can be easily done in polynomial time. 

\item Suppose that $s_j = q^{+}(C^i_j)$. Different from the previous case, some of the positive $x_{ia}$, where $a \in C^i_j \cap \LL^i_{\succ a^{\dagger}}$ will not contribute to  $x(A^i_j,  a^{\dagger})$. Our idea is to ``pin down'' the lowest ranking applicant $a^{\ddag} \in A^i_j$. 
After we pin down $a^{\ddag}$, we know that the positive $x_{ia}$s that contribute to  $x(A^i_j,  a^{\dagger})$ are those with $a \in C^i_j \cap \LL^i_{\succ a^{\ddag}}$, while those with 
$a \in C^i_j \cap (\LL^i_{\prec a^{\ddag}} \cap \LL^i_{\succ a^{\dagger}})$ do not. So what remains to be done is to find out the $s_j-1$ applicants $a$ in $C^i_j \cap \LL^i_{\succ a^{\ddag}}$ with the smallest values $x(T(i,a) \backslash \{(i,a)\})$. We then enumerate all possible $a^{\ddag} \in C^i_j \cap \LL^i_{\succ a^{\dagger}}$ and find out the applicant $a^{\ddag}$ that gives the smallest $Z(C^i_j, s_j, a^{\dagger})$. The whole process can be done in polynomial time. 

\end{itemize}

We now explain how to generalize to the case that $\C(i)$ is a laminar family. In the previous simplified case that $\C(i)$ is a partition, we critically collect the following information:

\begin{quote} Suppose that $A^i_j \subseteq C^i_j$ and $0 \leq |A^i_j| \leq q^{+}(C^i_j)$ and all applicants in $A^i_j$ rank at least as high as $a^{\ddag}$ and strictly higher than $a^{\dagger}$. What is the choice of $A^i_j$ so that 
$x(S(A^i_j), a^{\dagger})+ \sum_{a \in A^i_j} x(T(i,a)\backslash \{(i,a)\})$ is minimized? 
\end{quote}

The above question motivates the following definition. 

\begin{definition} Assume that $A^i_j \subseteq C^i_j$, $0 \leq |A^i_j| \leq q^{+}(C^i_j)$ and all applicants in $A^i_j$ rank at least as high as $a^{\ddag}$, who in turn, ranks higher than $a^{\dagger}$. Furthermore, let $x(A^i_j, a^{\dagger})$ inherit the definition as defined in Definition~\ref{def:so}. We define

$$Z(C^i_j, s_j, a^{\ddag}, a^{\dagger}) = min_{A^i_j: A^i_j \subseteq C^i_j, |A^i_j|= s_j, 
\mbox{all applicants in $A^i_j$ rank at least as high as $a^{\ddag}$.}} x(A^i_j, a^{\dagger}).$$

For the case that $Z(C^i_j, s_j, a^{\ddag}, a^{\dagger})$ is not well-defined, e.g., there is no feasible tuple of length $s_j$ so that all applicants in $A^i_j \subseteq C^i_j$ rank at least as high as $a^{\ddag}$, or $a^{\dagger} \succeq_{i} a^{\ddag}$, 
let $Z(C^i_j, s_j, a^{\ddag}, a^{\dagger})$ be an arbitrary large value. 

\end{definition}
In the following, we also associate each $Z(C^i_j, s_j, a^{\ddag}, a^{\dagger})$ with the corresponding feasible tuple $A^i_j$ that realizes this value\footnote{If $Z(C^i_j, s_j, a^{\ddag}, a^{\dagger})$ is some arbitrary large value, let its corresponding tuple be $\phi$ and indeed this will not matter.}. This helps us to identify the tuple that is violated in Constraint~(\ref{equ:lpThird}).

\begin{theorem} For all classes $C^i_j \in \C(i)$, we can correctly calculate all possible 
$Z(C^i_j, s_j, a^{\ddag}, a^{\dagger})$ in polynomial time, for all $a^{\dagger}$ and $a^{\ddag}$ combinations. 
\label{thm:so}
\end{theorem}

\begin{proof} We first remark that in the calculation, we indeed only need to find out all the values of $Z(C^i_j, s_j, a^{\ddag}, a^{\dagger})$ for $a^{\ddag} \in C^i_j$. For the case of $a^{\ddag} \not \in C^i_j$, simply copy the value of 
$Z(C^i_j, s_j, \overline{a^{\ddag}}, a^{\dagger})$ where $\overline{a^{\ddag}} \in C^i_j$ and is the lowest ranking applicant that ranks higher than $a^{\ddag}$. If there is no such applicant,  let $Z(C^i_j, s_j, a^{\ddag}, a^{\dagger})$ be an arbitrary large value. 

We now give an inductive proof based on the height of class $C^i_j$ in the tree structure of $\C(i)$. The base case is when $C^i_j$ is a leaf class. Then the calculation of all $Z(C^i_j, s_j, a^{\ddag}, a^{\dagger})$ can be done in essentially the same way as we have shown in the case that $\C(i)$ is a partition. 

For the induction step, let $C^i_j$ be a non-leaf class and assume that $a^{\ddag} \in C^i_{k'} \in c(C^i_j)$. To calculate $Z(C^i_j, s_j, a^{\ddag}, a^{\dagger})$, we need to find out a feasible tuple $A^i_j$ of size $s_j$, all of whose applicants rank at least as high as $a^{\ddag}$ so that 
$x(A^i_j, a^{\dagger})$ is minimized. 

Observe that a feasible tuple $A^i_j$ can be decomposed into a set of tuples $A^i_j = 
\bigcup_{C^i_k \in c(C^i_j)} A^i_k$, where 
$A^i_k \subseteq C^i_k \in c(C^i_j)$. 

\begin{enumerate}

\item Suppose that $s_j < q^{+}(C^i_j)$. Then by definition, $x(S(A^i_j), a^{\dagger}) = \sum_{C^i_k \in c(C^i_j)} x(S(A^i_k), a^{\dagger})$. So 
$$x(A^i_j, a^{\dagger}) = \sum_{C^i_k \in c(C^i_j)}\left[\sum_{a \in A^i_k}x(T(i,a)\backslash \{(i,a)\}) + x(S(A^i_k), a^{\dagger})\right].$$ 

For each class $C^i_k \in c(C^i_j)$, the minimum quantity $\sum_{a \in A^i_k}x(T(i,a)\backslash \{(i,a)\}) + x(S(A^i_k), a^{\dagger})$ is exactly $Z(C^i_k, s_k, a^{\ddag}, a^{\dagger})$. As a result, for each class $C^i_k \neq C^i_{k'}$, let $0 \leq s_k \leq q^{+}(C^i_k)$, and for class $C^i_{k'}$, let $0 \leq s_{k'}\leq q^{+}(C^i_{k'})-1$:  

$$Z(C^i_j, s_j, a^{\ddag}, a^{\dagger}) = x(T(i,a^{\ddag})) + 
min_{s_k: \sum_{s_k}= s_j-1}\sum_{C^i_k \in c(C^i_j)} Z(C^i_k, s_k, a^{\ddag}, a^{\dagger}).$$ 

Thus, we can find out $Z(C^i_j, s_j, a^{\ddag}, a^{\dagger})$ by dynamic programming. 

\item Suppose that $s_j = q^{+}(C^i_j)$. Note that this time since the class $C^i_j$ will be ``saturated'', the term $x(S(A^i_j), a^{\dagger})$ does not get any positive values $x_{ia}$, provided that $a \in C^i_j \cap (\LL^i_{\succ a^{\dagger}} \cap \LL^i_{\prec a^{\ddag}})$. So 
$x(S(A^i_j), a^{\dagger}) = \sum_{C^i_k \in c(C^i_j)} x(S(A^i_k), a^{\ddag})$ and this implies that

$$x(A^i_j, a^{\dagger}) = \sum_{C^i_k \in c(C^i_j)}\left[\sum_{a \in A^i_k}x(T(i,a)\backslash \{(i,a)\}) + x(S(A^i_k), a^{\ddag})\right].$$ 

\end{enumerate}

Let $\overline{a}^{\ddag}$ be the lowest ranking applicant that ranks higher than $a^{\ddag}$. Then for each class, $C^i_k \in c(C^i_j)$, the minimum quantity $\sum_{a \in A^i_k}x(T(i,a)\backslash \{(i,a)\}) + x(S(A^i_k), a^{\ddag})$ is exactly 
$Z(C^i_k, s_k, \overline{a^{\ddag}}, a^{\ddag})$. Assuming that 
for each class $C^i_k \neq C^i_{k'}$, let $0 \leq s_k \leq q^{+}(C^i_k)$, and let 
$0 \leq s_{k'}\leq q^{+}(C^i_{k'})-1$, we have 

$$Z(C^i_j, s_j, a^{\ddag}, a^{\dagger}) = x(T(i,a^{\ddag})) + 
min_{s_k: \sum_{s_k}= s_j-1}\sum_{C^i_k \in c(C^i_j)} Z(C^i_k, s_k, \overline{a^{\ddag}}, 
a^{\ddag}).$$ 

As before, this can be calculated by dynamic programming. \qed

\end{proof}

Now choose the smallest $Z(C^i_{\sharp}, Q(i)-1, a^{\ddag}, a^{\dagger})$ among all possible $a^{\ddag}$ who rank higher than $a^{\dagger}$ and assume that $A^i_{\sharp}$ is the corresponding tuple. It is easy to see that among all feasible tuples $A^i$ of length $Q(i)$ whose lowest ranking applicant is $a^{\dagger}$, the one has the smallest comb value  $x(K(i, S(A^i))$, is exactly the tuple $A^i_{\sharp} \cup \{a^{\dagger}\}$. 

\subsubsection*{Separating Constraint~(\ref{equ:lpFifth})}

We again make use of dynamic programming. The idea is similar to the previous one and the task is much simpler, so we will be brief. 

Suppose that we are checking all the class-tuples $\mathbb{T}^i_j$ corresponding to class $C^i_j$. Let $T^i_{j, a^{\dagger}} \subseteq \mathbb{T}^i_j$ be the subset of class-tuples whose lowest ranking applicants is $a^{\dagger}$. We need to find out the class-tuple 
$t^i_{j, a^{\dagger}} \in T^i_{j, a^{\dagger}}$ with the smallest value

$$ x(T(i,a^{\dagger})\backslash \{(i,a^{\dagger})\}) + \sum_{a \in t^i_{j, a^{\dagger}} \backslash \{a^{\dagger}\}} x(T(i,a)\backslash \{(i,a)\}), $$ 

\noindent and check whether this value is no less than $\sum_{a \in C^i_j \cap \LL^i_{\prec a^{\dagger}}}x_{ia}$. If it is, then we are sure that all class-tuples in $T^i_{j, a^{\dagger}}$ satisfy Constraint~(\ref{equ:lpFifth}), otherwise, we find a violated constraint. The above quantity 
can be easily calculated by dynamic programming as before. 

\section{A Counter Example for Section~\ref{sec:median}}

The example shown in Figure~\ref{fig:notMedian} contains five stable matchings. If we apply the median choice operation on all of them, we get the stable matching $\mu_2$, which does not give institutes $i_1$ and $i_2$ their lexicographical median outcome. 

\begin{figure*}[h]\footnotesize
\hrule
\begin{tabbing}
\hspace{0pt}\=\hspace{140pt}\=\hspace{180pt}\=
\hspace{160pt}\=\hspace{30pt}\=\hspace{15pt}\\

\> Institute Preferences \>  Classifications     \> Class Bounds   \\
\> $i_1\mbox{:} a_x a_y a_1 a_2 a_3 a_4   $  \> $C^1_1=\{a_1,a_2\}$, 
$C^1_2 = \{a_3,a_4\}$ \> $Q(i_1)=2$, $q^{+}(C^1_1)=1$, $q^{+}(C^1_2)=1$   \\
\> $i_2\mbox{:} a_z a_w a_2 a_1 a_4 a_3 $  \> $C^2_1 = \{a_1,a_2\}$, $C^2_2 = \{a_3,a_4\}$ \> $Q(i_2)=2$, $q^{+}(C^2_1)=1$, $q^{+}(C^2_2)=1$    \\
\> $i_3\mbox{:} a_1 a_2 a_3 a_4  a_x a_y a_z a_w$ \>  \>
 $Q(i_3)=4$ \\\\\\

\> Applicant Preferences \\
\> $a_1\mbox{:} i_2 i_1 i_3  $ \\
\> $a_2\mbox{:} i_1 i_2 i_3 $   \\
\> $a_3\mbox{:} i_2 i_1 i_3$   \\
\> $a_4\mbox{:} i_1 i_2 i_3$ \\
\> $a_x\mbox{:} i_3 i_1$ \\
\> $a_y\mbox{:} i_3 i_1$ \\
\> $a_z\mbox{:} i_3 i_2$ \\
\> $a_w\mbox{:} i_3 i_2$ \\\\\\

\> Stable Matchings \\
\> $\mu_{1} = \{(i_1; a_x, a_y), (i_2;a_z, a_w), (i_3;a_1, a_2, a_3, a_4)\}$ \\
\> $\mu_{2} = \{(i_1; a_1, a_3), (i_2;a_2, a_4), (i_3;a_x, a_y, a_z, a_w)\}$ \\
\> $\mu_{3} = \{(i_1; a_1, a_4), (i_2;a_2, a_3), (i_3;a_x, a_y, a_z, a_w)\}$ \\
\> $\mu_{4} = \{(i_1; a_2, a_3), (i_2;a_1, a_4), (i_3;a_x, a_y, a_z, a_w)\}$ \\
\> $\mu_{5} = \{(i_1; a_2, a_4), (i_2;a_1, a_3), (i_3;a_x, a_y, a_z, a_w)\}$ \\

\end{tabbing}
\vspace*{-0.35in}
\caption{An example of median choice stable matching which does not give the institutes their lexicographically median outcome.}
\vspace*{0.03in}
\hrule
\label{fig:notMedian}
\end{figure*}

\end{document}